\documentclass[pdflatex,sn-mathphys-num]{sn-jnl}

\usepackage{graphicx}%
\usepackage{multirow}%
\usepackage{amsmath,amssymb,amsfonts}%
\usepackage{array}
\usepackage{amsthm}%
\usepackage{mathrsfs}%
\usepackage[title]{appendix}%
\usepackage{xcolor}%
\usepackage{textcomp}%
\usepackage{manyfoot}%
\usepackage{booktabs}%
\usepackage{algorithm}%
\usepackage{algorithmicx}%
\usepackage{algpseudocode}%
\usepackage{listings}%
\usepackage{subcaption}
\usepackage{colortbl}

\theoremstyle{thmstyleone}%
\theoremstyle{thmstyletwo}%

\theoremstyle{thmstylethree}%

\begin{document}

\title[MlPET]{MlPET: A Localized Neural Network Approach for Probabilistic Post-Reconstruction PET Image Analysis Using Informed Priors}

\author*[1]{\fnm{Thomas Mejer} \sur{Hansen}}\email{tmeha@geo.au.dk}

\author[2]{\fnm{Nana} \sur{Christensen}}\email{nana.christensen@rm.dk}

\author[2,3]{\fnm{Mikkel} \sur{Vendelbo}}\email{mhve@biomed.au.dk}

\affil*[1]{\orgdiv{Department of Geoscience}, \orgname{Aarhus University}, \orgaddress{\street{H\o gh-Guldbergs Gade 2}, \city{Aarhus C}, \postcode{8000}, \country{Denmark}}}

\affil[2]{\orgdiv{Department}, \orgname{Organization}, \orgaddress{\street{Street}, \city{Aarhus C}, \postcode{8000},\country{Denmark}}}

\affil[3]{\orgdiv{Department}, \orgname{Organization}, \orgaddress{\street{Street}, \city{City}, \postcode{610101}, \state{State}, \country{Country}}}

\abstract{
    We present \textit{MlPET}, a fast localized machine learning approach for probabilistic PET image analysis that addresses noise and resolution limitations in conventional PET imaging. Our method transforms the traditional inverse problem into computationally efficient local analyses using a neural network that directly computes posterior statistics without expensive sampling methods. MlPET incorporates knowledge of the Point Spread Function, a spatially correlated noise model, and informed priors about tissue properties. We evaluate our approach on phantom data from three PET scanners (Siemens Vision Quadra, Siemens Biograph, and GE Discovery MI). Results demonstrate that MlPET achieves superior contrast recovery coefficients approaching 1.0 for all sphere sizes (including 10 mm spheres), substantially reduced noise, and approximately 3.5-fold improvement in spatial resolution compared to conventional reconstruction. Notably, MlPET at 40--80 seconds acquisition time outperforms conventional PET at 900 seconds, suggesting a potential 10--20 fold reduction in required scan time while maintaining or improving diagnostic quality. This approach shows promise for enhancing small lesion detection while enabling shorter scan times or reduced radiation exposure.
    }

\abstract{
\textbf{Purpose:} To develop and evaluate MlPET, a fast localized machine learning approach for probabilistic PET image analysis. The method addresses the trade-off between noise and spatial resolution in conventional reconstructions.

\textbf{Methods:} 
Building on a probabilistic deconvolution framework with informed priors, MlPET replaces computationally demanding Markov chain Monte Carlo sampling with a localized neural network trained to directly estimate the posterior mean of voxel activity from small image neighborhoods. The method incorporates scanner-specific point spread functions (PSF), spatially correlated noise modeling, and flexible prior information. Performance was evaluated on NEMA IEC phantom data acquired on three PET systems (GE Discovery MI, Siemens Biograph Vision 600, and Siemens Biograph Vision Quadra) under varying reconstruction settings and acquisition times.

\textbf{Results:} 
On NEMA IEC phantom data, MlPET produced contrast-recovery coefficients consistently higher than standard PET and frequently close to 1.0 (including the 10 mm sphere in several settings), while simultaneously reducing background noise and improving spatial definition. The effective point-spread function (PSF) full width at half maximum (FWHM) was on average reduced from about 2 mm in standard PET to below 1 mm with MlPET, corresponding to roughly a 2.5× decrease in effective blur. Comparable image quality was obtained at 40–80 s acquisition time using MlPET versus 900 s with conventional PET.

\textbf{Conclusions:} 
MlPET provides a computationally efficient approach for quantitative probabilistic post-reconstruction PET image analysis. By combining informed priors with the speed of a neural network, it achieves both noise suppression and resolution enhancement without altering reconstruction algorithms. The method shows promise for improved small-lesion detectability and quantitative reliability in clinical PET imaging. Future clinical studies will evaluate its performance on patient data and quantify effects under realistic prior uncertainty.
}


\maketitle

\begin{center}
\textit{Under review in EJNMMI Physics}
\end{center}

\section{Background}\label{sec1}

Positron Emission Tomography (PET) provides quantitative images of tracer uptake reflecting tissue metabolism and has become indispensable in oncology and functional imaging \citep{fletcher2008recommendations,bailey2005positron}. Modern PET systems combine high sensitivity, time-of-flight (TOF) capability, and advanced reconstruction algorithms, yet accurate identification of small regions of increased uptake remains challenging. Image noise and limited spatial resolution, mainly due to the finite number of detected counts and system blurring, constrain lesion detectability \citep{qi2003unified,panin2006fully,nuyts2014unconstrained,tong2010noise,stute2013practical}.

Efforts to address these limitations can be broadly divided into \textit{pre-reconstruction} and \textit{post-reconstruction} approaches.  
Pre-reconstruction methods incorporate physical and statistical models into iterative algorithms such as MLEM \citep{shepp1982maximum}, OSEM \citep{hudson1994accelerated}, or penalized likelihood frameworks \citep{teoh2015phantom}. Resolution modelling using the point-spread function (PSF) \citep{panin2006pet} and TOF information \citep{karp2008benefit} further improve image quality but do not fully eliminate the trade-off between noise and resolution.

Post-reconstruction approaches enhance already reconstructed PET images. Classical techniques include Gaussian and edge-preserving filtering \citep{hoffman1979quantitation,perona1990scale}, wavelet- and patch-based denoising \citep{dutta2013non}, and nonlocal means filtering, which reduce noise but at the cost of spatial resolution. More recently, deep learning has emerged as a data-driven alternative. Convolutional and encoder–decoder networks \citep{gong2018pet,schaefferkoetter2020convolutional}, as well as GAN-based methods \citep{wang20183d}, have demonstrated effective denoising while preserving details. Self-supervised and deep image prior schemes further reduce dependence on paired training data \citep{song2021noise2void,hashimoto2019dynamic}. Although powerful, such models are typically deterministic and do not quantify uncertainty—an important limitation for quantitative PET analysis. Moreover, methods that strongly suppress noise often reduce contrast, whereas those enhancing sharpness tend to amplify noise \citep{tohka2008deconvolution,gong2019pet}.

\subsection{Probabilistic Post-Reconstruction Analysis}

To address these limitations, \citet{hansen2023probabilistic} introduced a probabilistic framework for post-reconstruction PET analysis that integrates prior information, the PSF, and a correlated noise model into a single posterior probability distribution. Sampling this distribution using the extended Metropolis algorithm \citep{mosegaard1995monte} yields an ensemble of PET images, each consistent with all available information. From these samples, statistical quantities such as the posterior mean activity, variance, and the probability of elevated tracer uptake can be derived, enabling quantitative uncertainty analysis and improved lesion detectability.

However, the sampling approach is computationally demanding and currently limited to small image volumes.  
To address this, we build on the concept of \citet{hansen2022use}, who demonstrated in a geoscientific context that neural networks can be trained to directly estimate posterior statistics, without the need for explicit sampling. By adopting this idea for PET image analysis, we introduce a localized neural network that directly estimates the posterior mean of voxel activity from reconstructed PET data and informed priors.

This MlPET framework substantially reduces computational cost while preserving the statistical rigor of the probabilistic formulation. It enables fast, voxelwise analysis of post-reconstruction PET images and naturally produces mean activity maps with both reduced noise and improved spatial resolution. In this work, we demonstrate and validate MlPET primarily on phantom data to establish a best-case performance benchmark, and discuss its potential for future application to clinical datasets.

\section{Methods and Theory} \label{sec:methods}

\subsection{Probabilistic analysis of PET Images}

Let \(\mathbf{\Phi} = [\phi_1, \phi_2, \ldots, \phi_N]\) represent the underlying (true) activity concentration field at the time of scanning, and \(\mathbf{\Phi}_{\text{PET}} = [{\phi_{\text{PET}}}_1, {\phi_{\text{PET}}}_2, \ldots, {\phi_{\text{PET}}}_N]\) the corresponding reconstructed PET image. The relationship between the true, \(\mathbf{\Phi}\), and reconstructed activity fields, \(\mathbf{\Phi}_{\text{PET}}\), (the forward problem) can be expressed as:
\begin{equation}
    \mathbf{\Phi}_{\text{PET}} = g_{\text{PSF}}(\mathbf{\Phi}) + \text{n}(\mathbf{\Phi}),
    \label{eq:forward_nonlin}
\end{equation}
where \(g_{\text{PSF}}\) represents a convolution operator describing the system point-spread function (PSF), and \(\text{n}(\mathbf{\Phi})\) is the noise term. 
In the following, we assume a linear operator, such that $g_{\text{PSF}}(\mathbf{\Phi}) = \mathbf{G}_{\text{PSF}} \mathbf{\Phi}$.
Both \(\mathbf{G}_{\text{PSF}}\) and \(\text{n}(\mathbf{\Phi})\) depend on the PET system, reconstruction algorithm, and tracer, and hence differ between scanners.

We consider the \textit{inverse} problem of inferring information about \(\mathbf{\Phi}\) from \(\mathbf{\Phi}_{\text{PET}}\) in a probabilistic framework following \citet{tarantola2005inverse}, where the solution is the posterior probability density:
\begin{align}
    \sigma(\mathbf{\Phi}) \propto \rho(\mathbf{\Phi}) \cdot L(\mathbf{\Phi}),
    \label{eq:posterior}
\end{align}
with \(\rho(\mathbf{\Phi})\) denoting the prior and \(L(\mathbf{\Phi})\) the likelihood,
\begin{align}
    L(\mathbf{\Phi}) = f_d(\mathbf{\Phi}_{\text{obs}} - \mathbf{G}_{\text{PSF}} \mathbf{\Phi}),
    \label{eq:likelihood}
\end{align}
which measures how well the predicted image \(\mathbf{G}_{\text{PSF}}\mathbf{\Phi}\) explains the observed reconstruction \(\mathbf{\Phi}_{\text{obs}}\) within the assumed noise model.

This probabilistic framework, first demonstrated for PET by \citet{hansen2023probabilistic}, enables the explicit incorporation of complex prior information, even when no analytical expression for \(\rho(\mathbf{\Phi})\) exists. Sampling from the posterior is performed using the extended Metropolis algorithm \citep{mosegaard1995monte}, a variant of the Metropolis--Hastings method \citep{metropolis1949mcm}, which only requires an algorithm capable of generating realizations from the prior and evaluating the likelihood.
The approach can in principle, tough computationally demanding, be applied to smaller local susbsets of a 3D PET image, but becomes computationally intractable for full 3D PET images (Typically with \(>25\) million voxels). To overcome this limitation, we introduce a \textit{localized inversion} strategy.

\subsubsection{Localized inversion}
\begin{figure}[h]
    \centering
    \includegraphics[width=0.75\textwidth]{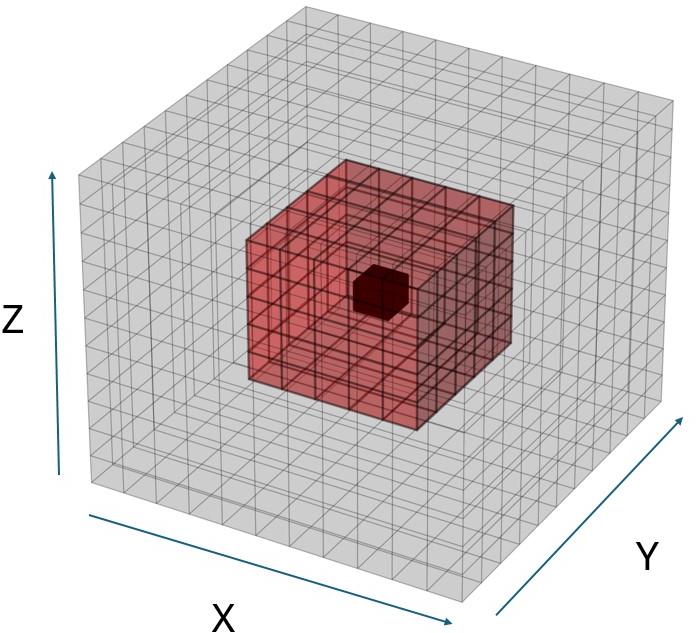}
    \caption{Schematic of a localized neighborhood $\mathbf{\Phi}_l$ (red voxels) centered on the voxel of interest $\phi_{ic}$ (black). Gray voxels indicate the full model domain $\mathbf{\Phi}$.}
    \label{fig:2d3dwireframe}
\end{figure}

The probabilistic problem in Eq.~\ref{eq:posterior} can be reformulated locally for a neighborhood around each voxel, allowing posterior estimation through a series of smaller, tractable subproblems.  
For a given central voxel \(\phi_{ic}\), the local data consist of observed PET values \(\mathbf{\Phi}_{\text{PET},l}\) within a cubic neighborhood defined by widths \(w_x, w_y, w_z\), resulting in \(N_l = (2w_x+1)(2w_y+1)(2w_z+1)\) voxels.  
The corresponding local posterior distribution is then
\begin{align}
    \sigma(\mathbf{\Phi}_l) \propto \rho(\mathbf{\Phi}_l) \cdot L(\mathbf{\Phi}_l),
    \label{eq:posteriorlocal}
\end{align}
and the main quantity of interest is the marginal posterior of the central voxel, \(\sigma(\phi_{ic})\).  
This localized approach, originally developed and described in detail for geophysical electromagnetic data by \citet{hansen2021efficient}, treats each voxel neighborhood as an independent low-dimensional probabilistic inversion problem.  
Figure~\ref{fig:2d3dwireframe} illustrates the localized setup used in the present work.

Although each subproblem is low-dimensional, sampling Eq.~\ref{eq:posteriorlocal} using the extended Metropolis algorithm for all neighborhoods in a 3D image remains computationally intractable.  
To address this, we consider two alternatives: (1) the extended rejection sampler, used here as a reference method, and (2) a neural-network–based estimator that directly predicts local posterior statistics.

\paragraph{The extended rejection sampler}

The extended rejection sampler \citep{hansen2021efficient} generates realizations \(\mathbf{\Phi}_l^*\) from the prior and accepts them as posterior samples with probability
\begin{equation}
    P_{\text{acc}} = \frac{L(\mathbf{\Phi}_l^*)}{\max(L(\mathbf{\Phi}_l))}.
\end{equation}
A lookup table can be precomputed as
\begin{equation}
    \mathbf{T} = [\mathbf{\Phi}_l^*, \mathbf{\Phi}_{\text{PET},l}^*],
    \label{eq:lookup1}
\end{equation}
linking prior realizations and their corresponding forward responses.  
This removes the need for repeated forward simulations and enables fast, parallel evaluation of the likelihood function.  
However, exhaustive sampling across all neighborhoods in a full 3D volume remains computationally expensive, motivating the neural-network approach introduced below.

\subsubsection{Localized inversion using a neural network}

\citet{hansen2022use} proposed a method to estimate posterior statistics directly using a neural network (NN), avoiding explicit sampling of the posterior distribution.  
Training data for the NN are similar to the lookup table above but include additional simulated noisy data, $\mathbf{\Phi}_{\text{PET},l,\text{noise}}^*$, by adding a realization of the noise model to the noise-free forward response:
\begin{equation}
    \mathbf{T} = [\mathbf{\Phi}_l^*, \mathbf{\Phi}_{\text{PET},l}^*, \mathbf{\Phi}_{\text{PET},l,\text{noise}}^*].
    \label{eq:lookup2}
\end{equation}
The NN learns the mapping from noisy data \(\mathbf{\Phi}_{\text{PET},l,\text{noise}}^*\) to statistical descriptors \(F(\phi_{ic})\) of the posterior distribution for the central voxel,
\begin{equation}
    \mathbf{\Phi}_{\text{PET},l,\text{noise}}^* \mapsto F(\phi_{ic}),
\end{equation}
where \(F\) may represent the posterior mean, standard deviation, or other statistical properties.

\begin{figure}
    \includegraphics[width=\textwidth]{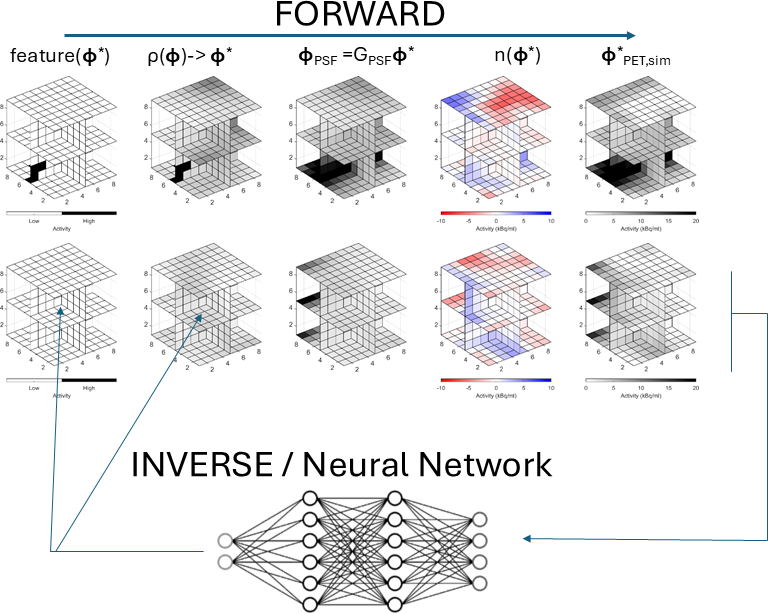}
    \caption{Example of a fully connected NN architecture used to estimate the posterior mean and standard deviation of the marginal posterior distribution of the central voxel $\phi_{ic}$ in a neighborhood $\mathbf{\Phi}_l$.}
    \label{fig:mlpet_nn}
\end{figure}

The NN architecture is fully connected, with an input layer of \(N_l\) neurons (one per voxel in the local neighborhood) and an output layer corresponding to the number of posterior statistics estimated.  
The loss function is the negative log-likelihood of the assumed marginal distribution (e.g., Gaussian). When only the mean is estimated, as will be used below, the mean squared error (MSE) is used.  
Figure~\ref{fig:mlpet_nn} illustrates the NN architecture for estimating the posterior mean and standard deviation.

\paragraph{Choice of neighborhood size}
Smaller neighborhoods reduce computational cost but can limit achievable resolution, whereas excessively large ones increase training time without additional benefit.
The neighborhood must be sufficiently large to encompass the effective PSF and the spatial correlation of the noise \citep{wilson1993noise,tong2010noise}.
Accordingly, the minimal neighborhood size that covers both the PSF and noise correlation length in all dimensions was selected.

\subsection{PET scanners and phantom data}
The MlPET method was tested on NEMA IEC body phantom data acquired on three PET systems: GE Discovery MI, Siemens Biograph Vision 600, and Siemens Biograph Vision Quadra Edge.  
The phantom contains six spherical inserts (10–37 mm diameter) with known activity concentrations.  
For GE Discovery, images were reconstructed both with and without PSF modeling; Vision 600 and Quadra data were reconstructed with OSEM (TOF + PSF modeling, 4 iterations, 5 subsets).  
For the Quadra system, acquisition times ranged from 1 s to 900 s (leading to 11 reconstructed phantom data sets).  
All reconstructions included standard corrections (attenuation, scatter, randoms, decay, and normalization) and Gaussian post-filtering (2–4 mm).

\subsubsection{Estimation of the noise}
\label{sec:noise}
\begin{figure}
    \centering
    \includegraphics[width=.45\textwidth]{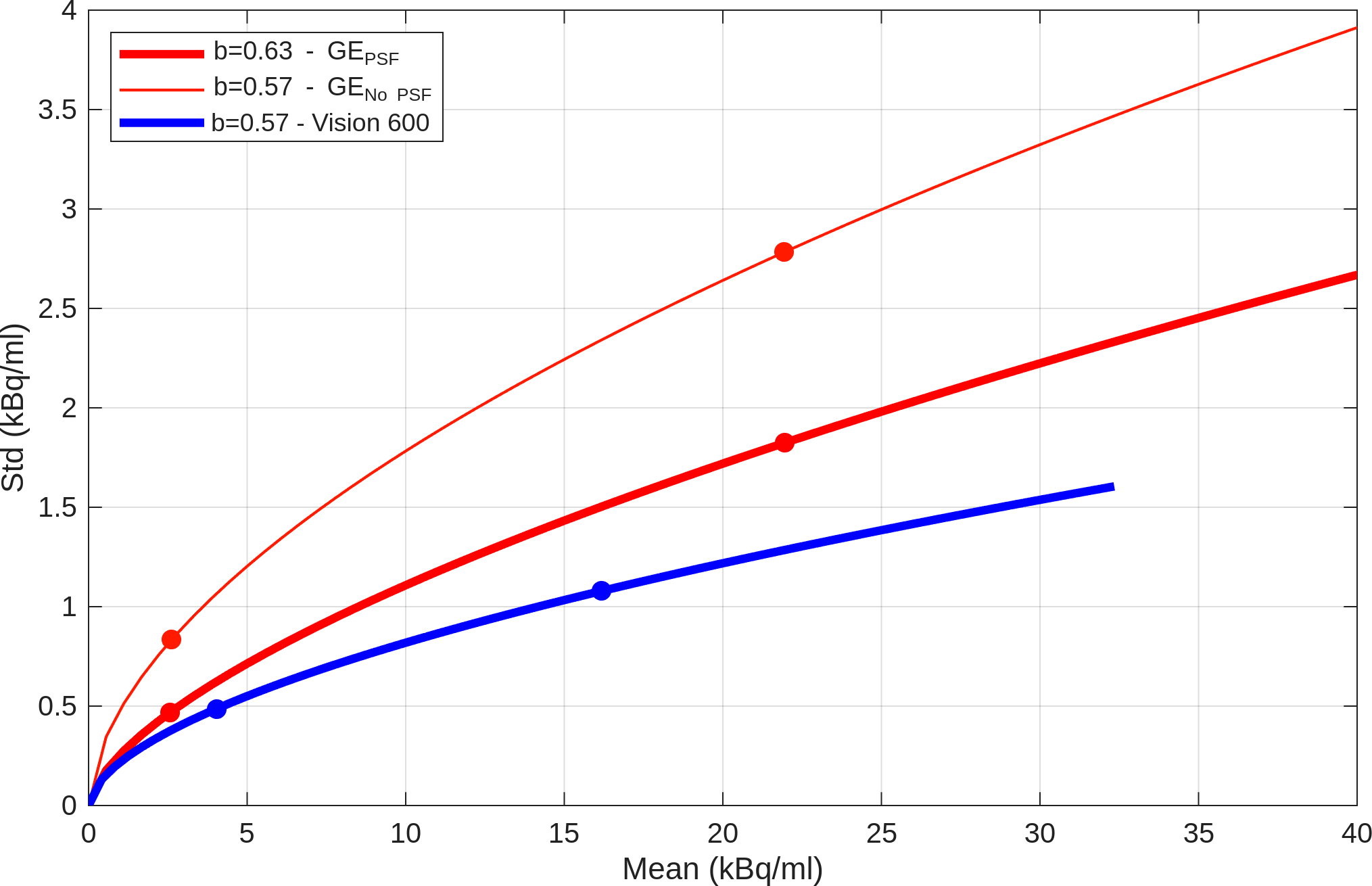}
    \includegraphics[width=.45\textwidth]{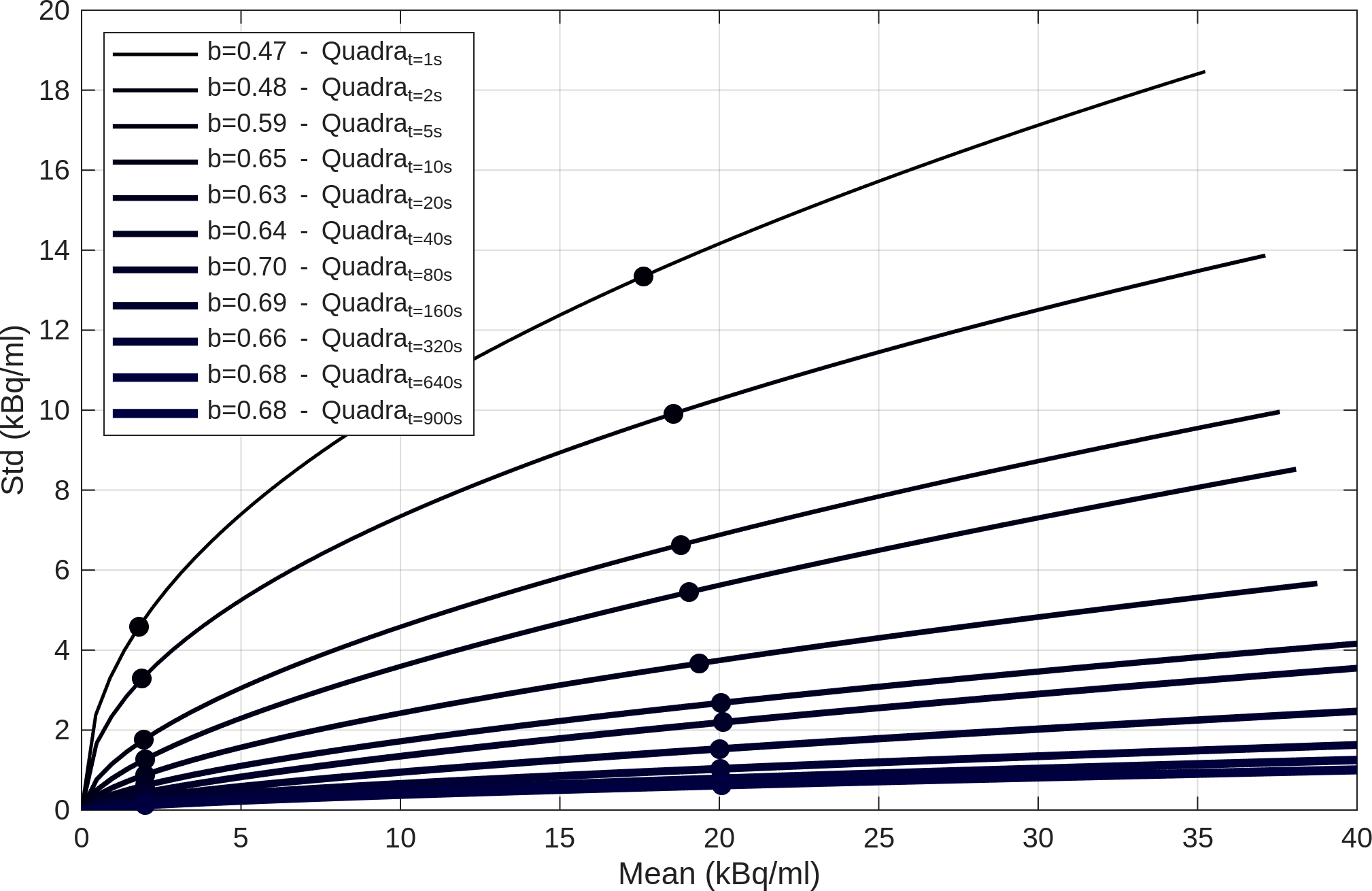}
    \caption{Estimated power-law relation between \(\phi_{\text{std}}\) and \(\phi_{\text{mean}}\) for GE Discovery and Vision 600 (left) and Vision Quadra Edge (right). Dots show measured pairs; lines show fitted power-law models (Eq.~\ref{eq:powerlaw}).}
    \label{fig:noise_ge_siemens}
\end{figure}

Noise is modeled as a stationary 3D multivariate Gaussian field with spatial correlation. 
The correlation length (expressed as the FWHM of a Gaussian-type covariance model) is estimated via semivariogram analysis of uniform background regions \citep{Goovaerts:1997}.  
The dependence of noise variance on mean activity is described by a power-law relation,
\begin{align}
    \phi_{\text{std}} = a \, \phi_{\text{mean}}^b,
    \label{eq:powerlaw}
\end{align}
where \(a\) is a scaling factor and \(b\) the power-law exponent. 
\(a\) and \(b\) are estimated by measuring local mean and standard deviation pairs \((\phi_{\text{mean}}, \phi_{\text{std}})\) within low and high activity regions in the use phantom data. 
Figure~\ref{fig:noise_ge_siemens} shows the estimated powerlaw relations for all considered  phantom data.

\subsubsection{Estimation of the effective PSF}
\label{sec:infer_psf}
We assume that the effective PSF can be represented by an isotropic Gaussian kernel characterized by its full width at half maximum (FWHM).  
The FWHM of the effective PSF, \(\text{PSF}_{\text{fwhm}}\), is estimated by minimizing the difference between the reconstructed PET image and the noise-free forward model of a reference phantom, as suggest in \cite{hansen2023probabilistic}:
\begin{equation}
    \text{LOSS}(\text{PSF}_{\text{fwhm}}) = \|\mathbf{\Phi}_{\text{PET}} - \mathbf{G}_{\text{PSF}} \mathbf{\Phi}_{\text{phantom}}\|.
\end{equation}
The kernel is assumed isotropic, with equal FWHM in all directions. 

Table \ref{tab:inferred_noise_params} summarizes the inferred noise parameters, as well as inferred the effective PSF, for all considered systems and phantoms.

\subsubsection{The prior model} 
\label{sec:prior}

The prior probability density function (pdf) \(\rho(\mathbf{\Phi})\) expresses prior beliefs about the underlying activity distribution \(\mathbf{\Phi}\), independent of the PET scanner, reconstruction algorithm, or noise model. 
In the present probabilistic framework, the prior provides an explicit way to incorporate medical or physical knowledge into the inversion process. 
Unlike implicit regularization methods, this approach requires a deliberate and transparent choice of prior model, which can then be visualized, validated, and discussed with clinical experts.

It is essential that the prior is constructed from biological or physical knowledge, not from reconstructed PET images, as these are already affected by blurring and noise. 
For medical imaging, prior information may come from clinical experience, tracer physiology, or known uptake patterns in specific tissues. 
For example, uptake of \textsuperscript{18}F-FDG is typically elevated in malignant or inflamed tissues due to increased glucose metabolism \citep{ben200918f}. 
Pathological regions such as malignant lymph nodes often exhibit sharp boundaries in the in vivo activity distribution, reflecting abrupt transitions between high- and low-uptake tissues. 
Such features should be captured in the prior model, even if they are blurred in the reconstructed PET images \citep{hoffman1979quantitation}.

No analytical expression for \(\rho(\mathbf{\Phi})\) is required; only an algorithm capable of generating realizations of the prior. 
This allows explicit, flexible construction of priors that reflect realistic spatial patterns and variability in tracer uptake. 
Below, two representative priors are defined: one for controlled phantom data, and one for clinical breast cancer imaging.

\paragraph{Prior model for NEMA phantom data}

The NEMA IEC body phantom simulates typical tracer distributions encountered in oncology, with spherical inserts of higher activity embedded in a uniform background. 
The corresponding prior, \(\rho_p(\mathbf{\Phi})\), is designed to reflect this known structure and to test the sensitivity of the inversion to an informed prior.

We assume three main activity domains:
\begin{enumerate}
    \item[(a)] a background region with near-zero activity,
    \item[(b)] a low-activity outer volume representing normal tissue, and
    \item[(c)] discrete high-activity spheres representing lesions.
\end{enumerate}

The activity levels for the low- and high-activity regions are drawn from uniform distributions within \(\pm 5\%\) of the nominal phantom values. 
This narrow variability defines a well-informed prior, challenging the inversion to reconcile the prior with measured data through the noise and PSF models. 
An informative prior can be challenging if and when either the forward and/or noise models are inaccurate, as will be discussed.
Such informed priors are useful for demonstrating both the potential benefits and the sensitivity of the probabilistic formulation.

\begin{figure}[h]
    \centering
    \includegraphics[width=\textwidth]{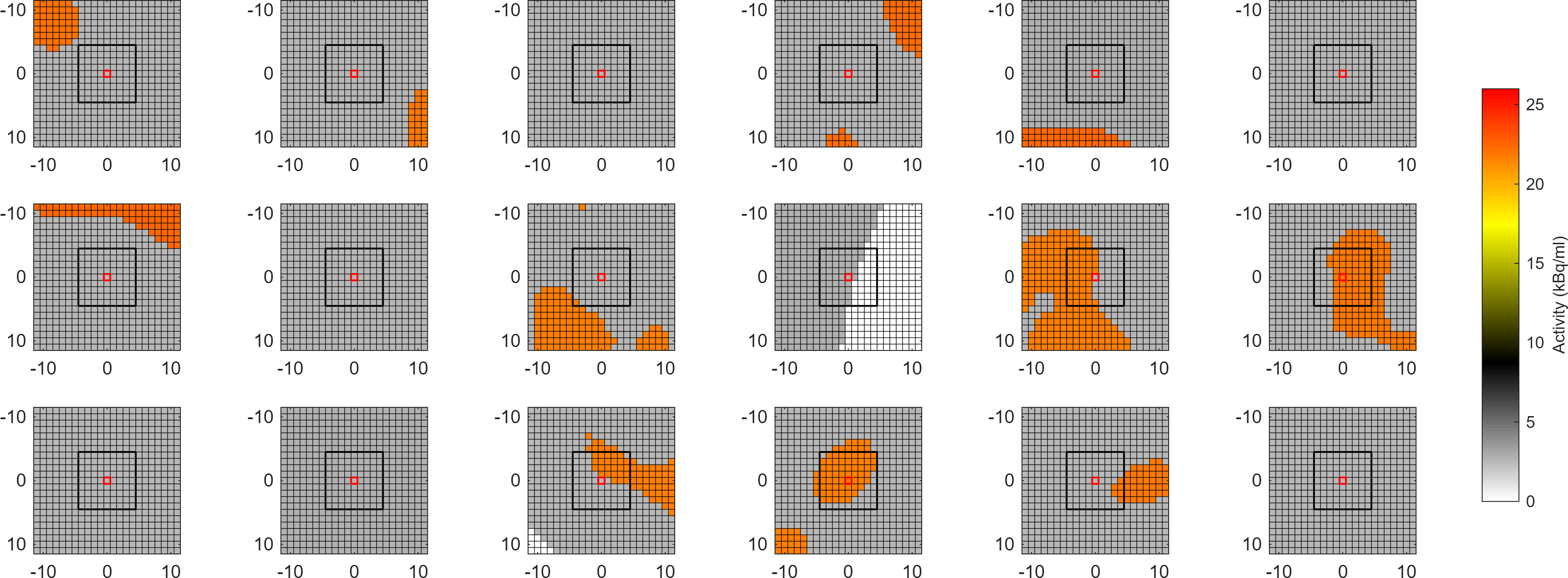}
    \caption{Example realizations from the NEMA phantom prior model $\rho_p(\mathbf{\Phi})$. 
    The black square marks the analyzed neighborhood, and the red square marks the central voxel.}
    \label{fig:priormodelA}    
\end{figure}

Overall, \(\rho_p(\mathbf{\Phi})\) captures three distinct classes—background, normal tissue, and high-uptake lesions—with sharp spatial boundaries between them. 
Figure~\ref{fig:priormodelA} shows example realizations of \(\rho_p(\mathbf{\Phi})\).

\paragraph{Lesion-based prior model for oncological PET imaging}

For clinical applications, a biologically informed prior, \(\rho_c(\mathbf{\Phi})\), was designed to represent the expected tracer uptake in breast cancer patients imaged with \textsuperscript{18}F-FDG PET. 
This prior reflects both physiological and pathological uptake patterns and provides a realistic foundation for probabilistic analysis of clinical data.

The following assumptions underpin \(\rho_c(\mathbf{\Phi})\):
\begin{enumerate}
    \item[(a)] Normal breast tissue exhibits low, spatially smooth FDG uptake.
    \item[(b)] The liver serves as a stable reference region for normalization.
    \item[(c)] Malignant lesions display elevated uptake with sharp boundaries relative to surrounding tissue.
    \item[(d)] Small metastatic nodes may appear as isolated high-uptake regions.
\end{enumerate}

Activity levels are modeled as spatially heterogeneous but structured. 
Normal tissue voxels are drawn from a uniform distribution centered on the patient-specific liver mean activity (with a standard devation of 10\% of the mean), while lesion regions are assigned activities 2–10 times higher, with sharp gradients at the boundaries. 
This reflects the physiological contrast between malignant and healthy tissues observed in FDG-PET.

Realizations of \(\rho_c(\mathbf{\Phi})\) are generated using a hierarchical stochastic model. 
First, a 3D multivariate Gaussian field describes the smooth background uptake, scaled to the patient’s liver activity distribution. 
Localized high-uptake regions are then introduced using truncated Gaussian fields that define discrete, spatially correlated lesions with variable shapes and intensities. 
The combination produces realistic tracer distributions consistent with expected clinical variability in FDG uptake.

\begin{figure}[h]
    \centering
    \includegraphics[width=\textwidth]{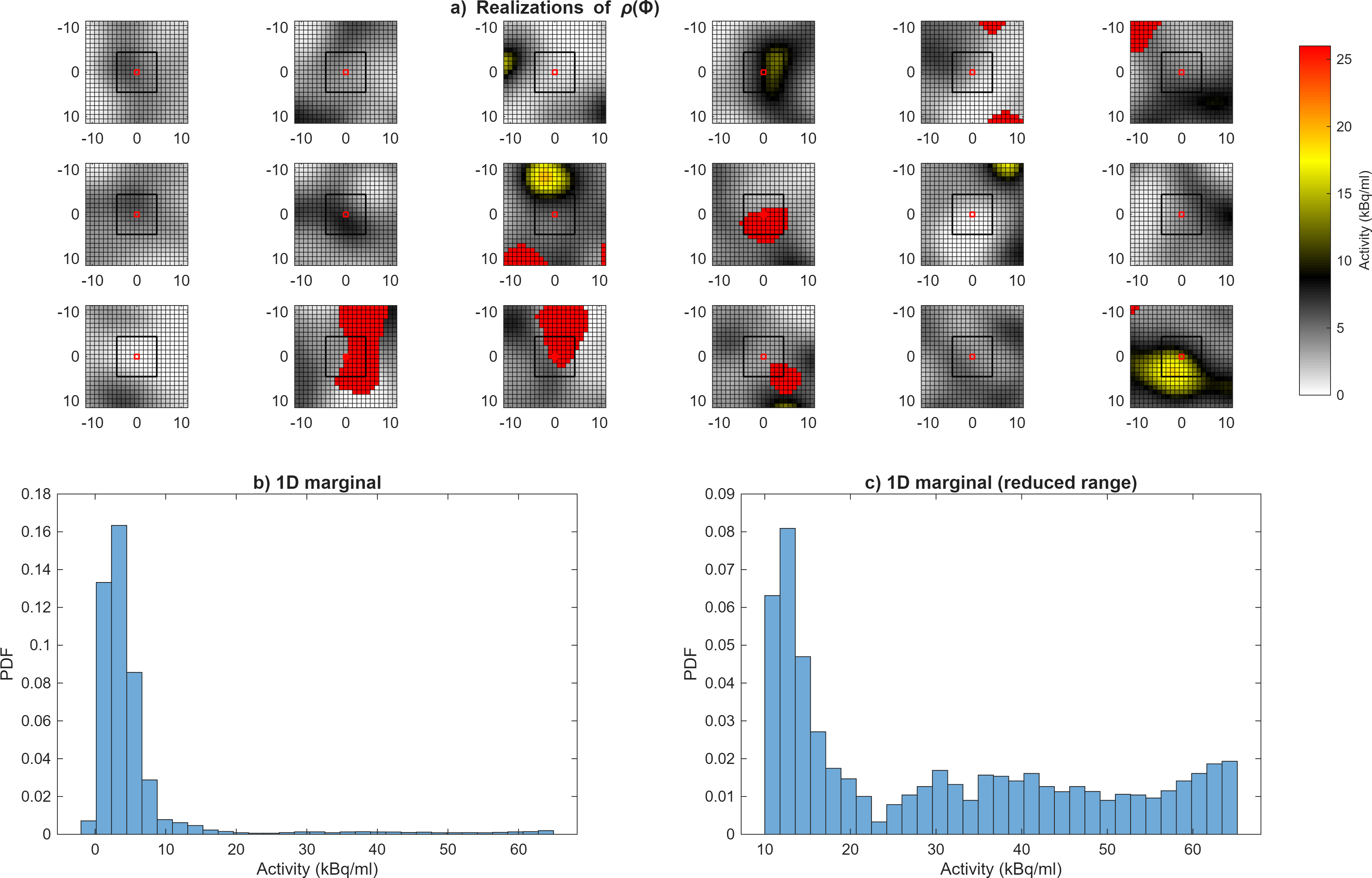}
    \caption{Example realizations from the clinical prior model $\rho_c(\mathbf{\Phi})$ representing expected activity distributions in breast cancer patients. 
    (a) Realizations from $\rho_c(\mathbf{\Phi})$. 
    (b) Full marginal prior distribution of the central voxel (black square in a). 
    (c) Marginal distribution for activity values above 10~kBq/ml for a patient with a liver mean of 10~kBq/ml and standard deviation of 2~kBq/ml.}
    \label{fig:priormodelC}
\end{figure}

Figure~\ref{fig:priormodelC} shows examples of realizations from \(\rho_c(\mathbf{\Phi})\), along with marginal prior distributions of the central voxel for two example patients with different liver activity reference levels.

These explicit prior formulations demonstrate how biological and physical knowledge can be encoded directly into probabilistic PET analysis. 
Different priors can easily be tailored to other tracers, organs, or diseases, allowing flexible integration of domain-specific information into quantitative PET interpretation.

\subsection{Evaluation metrics}
Quantitative performance was evaluated using standard NEMA metrics—contrast recovery coefficient (CRC) and background variability (expressed here as the coefficient of variation, COV)—together with additional measures of image quality including the contrast-to-noise ratio (CNR) and the effective PSF full width at half maximum (FWHM).

\paragraph{Contrast recovery coefficient (CRC):}
\begin{equation}
\text{CRC} = \frac{\text{Activity}_{\text{median}}^{\text{sphere}} / \text{Activity}_{\text{mean}}^{\text{background}}}{\text{True sphere activity} / \text{True background activity}}.
\label{eqn:CRC}
\end{equation}

\paragraph{Coefficient of variation (COV):}
\begin{equation}
\text{COV} = \frac{\text{SD}_{\text{background}}}{\text{Mean}_{\text{background}}} \times 100\%.
\label{eqn:COV}
\end{equation}

\paragraph{Contrast-to-noise ratio (CNR):}
\begin{equation}
\text{CNR} = \frac{\text{Activity}_{\text{median}}^{\text{sphere}} - \text{Activity}_{\text{mean}}^{\text{background}}}{\text{SD}_{\text{background}}}.
\label{eqn:CNR}
\end{equation}

\section{Results}\label{sec:results}
In the following, PET images acquired on three different scanner systems (the Siemens Biograph Vision 600, the GE Discovery MI, and the Siemens Biograph Vision Quadra) are analyzed using the localized probabilistic inversion method MlPET described in Section~\ref{sec:methods}.

The Biograph Vision 600 data are used to evaluate the sensitivity of MlPET to the choice of PSF range and the amplitude of the noise model (Section~\ref{sec:results_600}). 
The GE Discovery data are used to analyze the impact of applying MlPET to PET images reconstructed with and without PSF modeling (Section~\ref{sec:results_ge}). 
The Biograph Vision Quadra data are used to assess the effect of acquisition time on image quality and noise characteristics (Section~\ref{sec:results_quadra}). 
For all phantom-based experiments, the recovered activity distributions are compared directly to the known ground truth, \(\mathbf{\Phi}_{\text{Phantom}}\).

Finally, we demonstrate the applicability of MlPET to clinical data using an example from a breast cancer patient imaged with \textsuperscript{18}F-FDG PET/CT.

\subsection{LoPET vs MlPET}
Regardless of the chosen neighborhood size, the local posterior statistics \(F(\phi_{ic})\) should be identical when obtained using the extended Metropolis algorithm, the rejection sampler, or the NN-based approach, differing only in computational efficiency. Therefore, we compare the results from the rejection sampler and the neural network to verify that the NN has been trained correctly and can directly estimate the statistics of the marginal posterior distribution.

\textit{To come .......... if we do not include this we can shorten some of the text above about the localized rejection sampler.}

\subsection{Siemens Biograph Vision 600 – Sensitivity analysis}
\label{sec:results_600}
It is essential for any probabilistic inversion method that the noise model accurately represents the actual noise in the data, and that the chosen prior model is consistent with the true forward process. 
Even when care has been taken to estimate the noise characteristics and the PSF kernel, as described in Section~\ref{sec:noise}, some uncertainty in these estimates will always remain. 
For instance, it is well known that running the OSEM algorithm for many iterations can affect image quality \citep{barrett1994noise,wilson1994noise,tong2010noise}. 
In particular, the number of iterations influences the statistical properties of the noise and may introduce artifacts such as undershooting or overshooting near sharp transitions in activity.

To evaluate the robustness of MlPET with respect to these uncertainties, we investigate the sensitivity of the results to both the amplitude of the noise model and the width of the PSF kernel used in the forward model (Eq.~\ref{eq:forward_nonlin}). 
MlPET is applied using five different noise models, where the noise amplitude is scaled to \( NL = [5\%, 50\%, 100\%, 150\%, 200\%] \) of the estimated noise level. 
This allows assessment of both underestimation and overestimation of the noise variance. 
Similarly, five PSF kernel widths are considered, corresponding to \(PSF_r = [50\%, 75\%, 100\%, 125\%, 200\%]\) of the estimated PSF range. 
In total, 25 combinations of noise level and PSF range are analyzed to examine the sensitivity and stability of the probabilistic inversion process.

\subsubsection{Visual assessment}
\begin{figure}[h]
    \centering
    \includegraphics[width=1.00\textwidth]{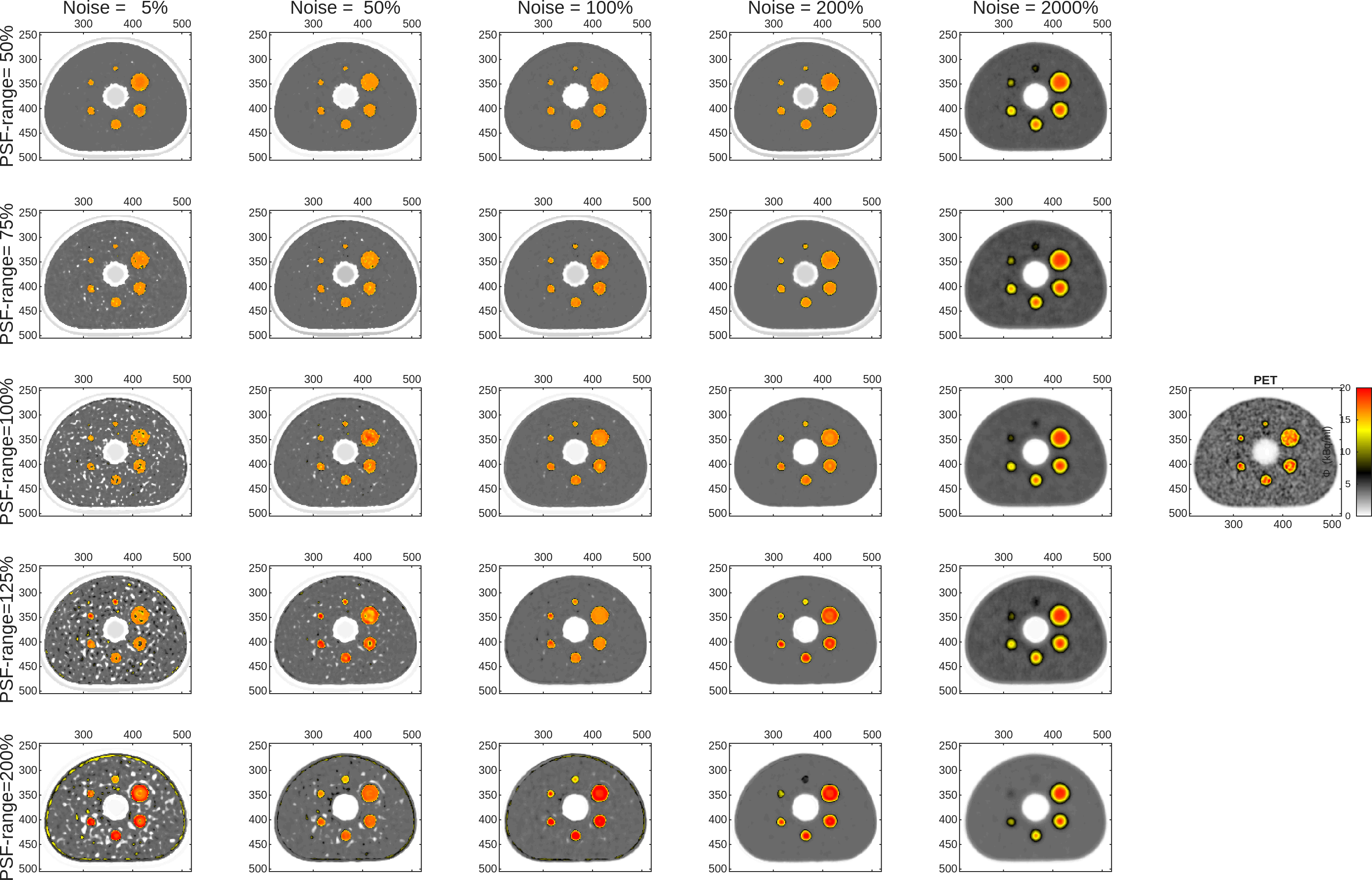}
    \caption{2D slice of the activity concentration obtained MlPET(bottom) (with different assumed noise level and PSF ranges) and PET (right) through the center of the 6 phantom spheres.}\label{fig:siemens_compare}
    \label{fig:siemens_visual}
\end{figure}

Figure~\ref{fig:siemens_visual} presents a visual comparison of 2D transaxial slices through the central plane of the NEMA phantom, showing the estimated posterior mean activity concentration obtained with different combinations of noise level and PSF range.

As expected, assuming an excessively high noise amplitude results in loss of spatial resolution, whereas assuming a too low noise amplitude causes overfitting of noise, producing artificially noisy images. 
High assumed noise levels do not introduce artifacts but cause the spheres to appear blurred. 
In contrast, underestimating the noise level leads to spurious fine structures that are not present in the true activity distribution, as the algorithm begins to fit random noise as data. 
To avoid such overfitting, and acknowledging that the 3D correlated Gaussian noise model may not perfectly represent the true noise characteristics, it is generally preferable to slightly overestimate the noise level when applying probabilstic inversion.

Similarly, overestimating the PSF range increases apparent image noise and blurring. 
The combination of an underestimated noise level and an overestimated PSF range produces the poorest results, characterized by overfitting.

\subsubsection{Quantitative analysis}
The qualitative observations are supported by a quantitative evaluation of noise characteristics and contrast recovery, as described in Section~\ref{sec:results}. 

\begin{figure}[h]
    \centering    
    \includegraphics[width=0.75\textwidth]{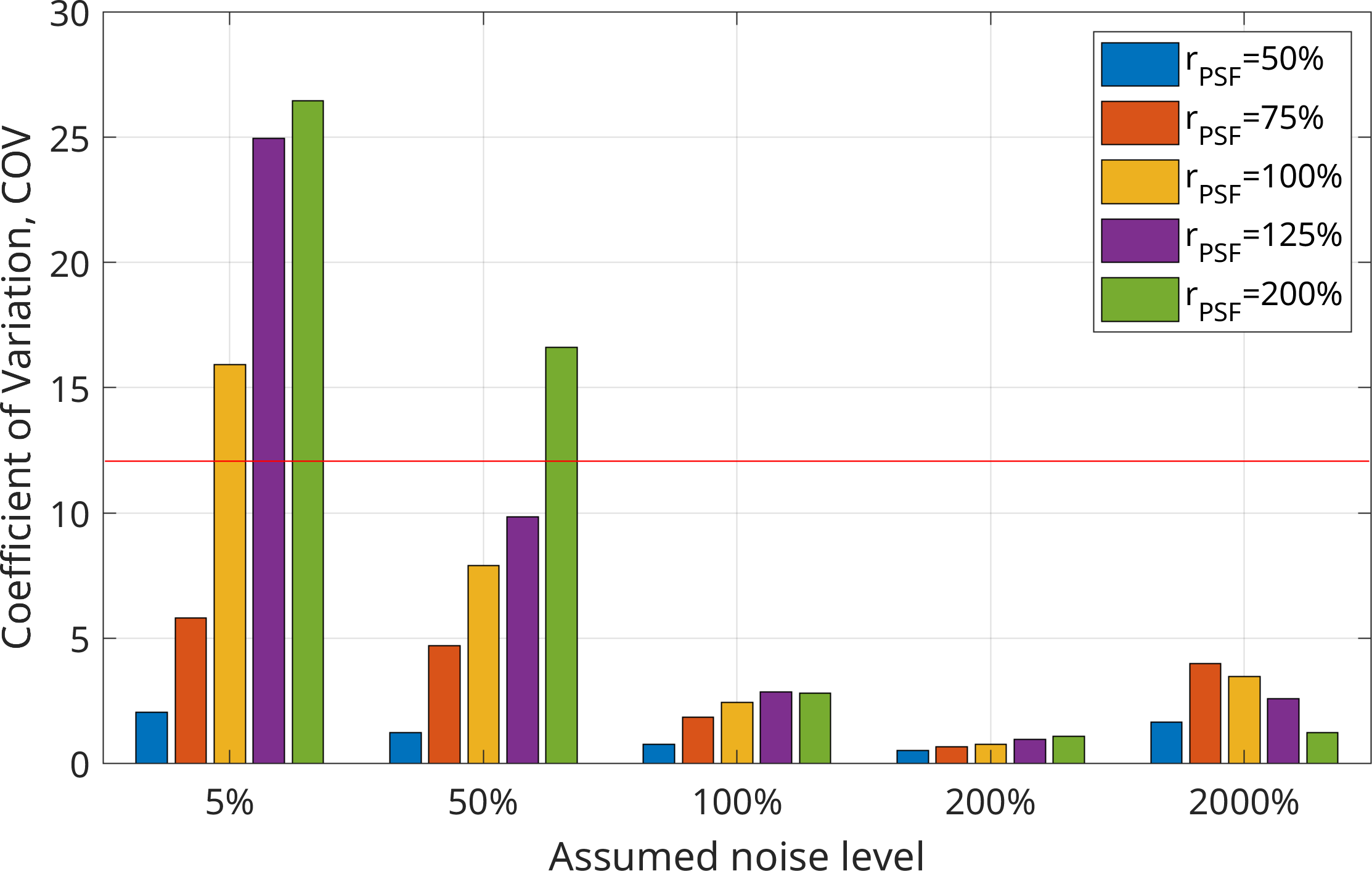}
    \caption{Coefficient of Variation (COV, Eqn. \ref{eqn:COV}) for difference noise levels and effective PSF range. Red line represents the COV obtained using the standard PET reconstruction.}
    \label{fig:siemens_cov}
\end{figure}

Figure~\ref{fig:siemens_cov} shows the Coefficient of Variation (COV, Eq.~\ref{eqn:COV}) for all 25 tested combinations of noise level and PSF range. 
As expected, underestimating the noise amplitude results in high COV values, indicating increased image noise. 
Conversely, overestimating the noise level reduces the COV but also leads to loss of spatial resolution, consistent with the visual impressions in Figure~\ref{fig:siemens_visual}.

\begin{figure}[h]
    \centering
    \includegraphics[width=0.75\textwidth]{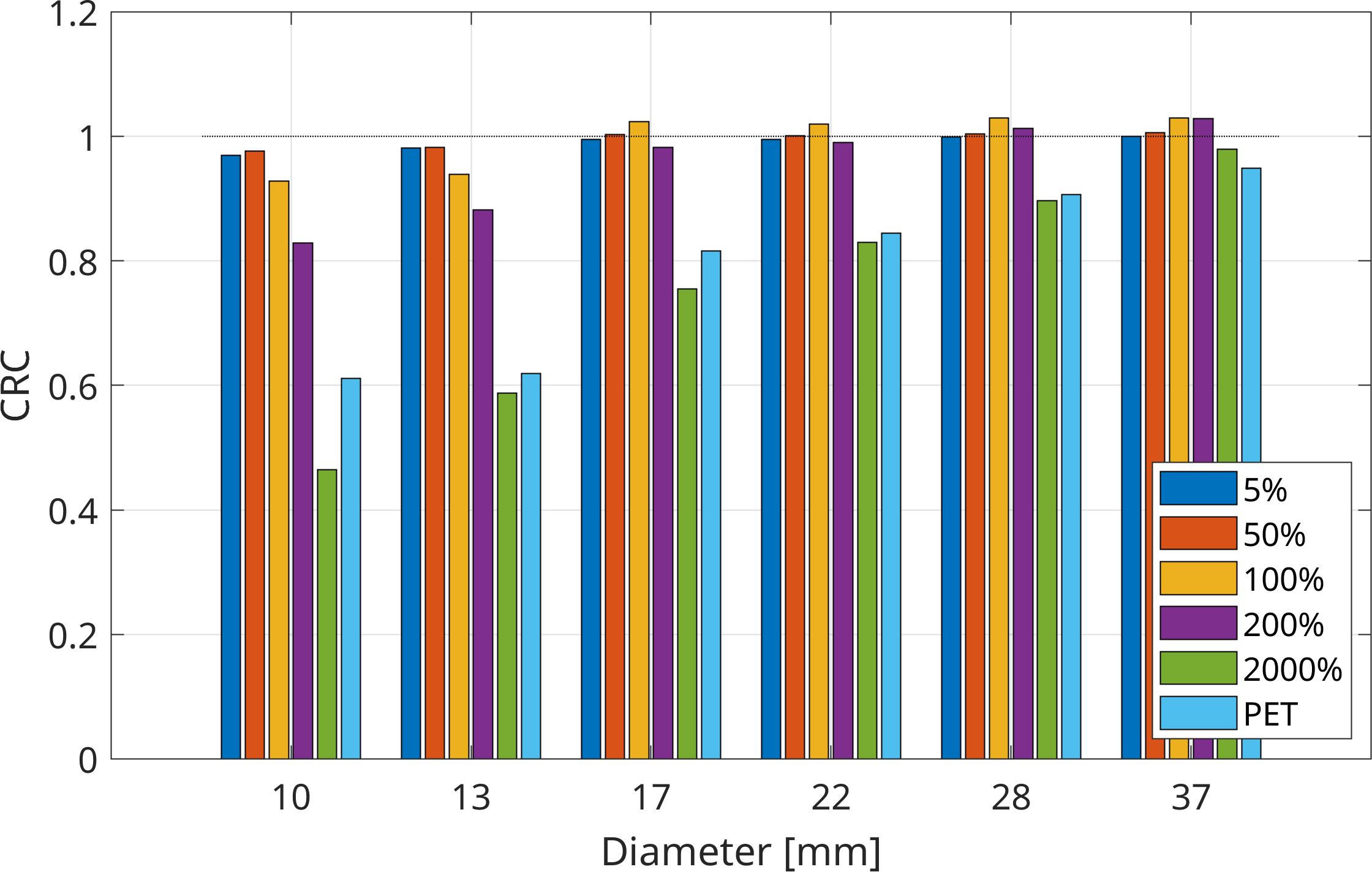}
    \caption{Contrast recovery coefficient (CRC, Eqn. \ref{eqn:CRC},) for the 6 spheres with varying noise amplitude \(NA= [5\%, 50\%, 100\%, 150\%, 2000\%]\)}
    \label{fig:siemens_crc}
\end{figure}

Figure~\ref{fig:siemens_crc} shows the Contrast Recovery Coefficient (CRC, Eq.~\ref{eqn:CRC}) for the six NEMA spheres under varying assumptions about the noise amplitude. 
MlPET yields CRC values consistently closer to one than the PET reference image for all sphere sizes and all noise amplitudes, except when the assumed noise level is extremely high (\(NL = 2000\%\)), in which case the smallest sphere (10\,mm diameter) becomes visibly blurred.

Overall, the sensitivity analysis demonstrates that MlPET achieves both noise reduction (Figure~\ref{fig:siemens_cov}) and improved resolution, as quantified by the CRC (Figure~\ref{fig:siemens_crc}), relative to standard PET reconstruction when realistic estimates of the effective PSF and noise model are used. 
The results also indicate that the noise amplitude should not be underestimated. 
Using a slightly higher noise level than the estimated value provides robustness to uncertainties in both the PSF and noise characterization. 
This choice results in a minor loss of resolution but effectively prevents overfitting, while still yielding markedly better noise–resolution performance than conventional reconstruction.

In contrast, simultaneous underestimation of the noise level and overestimation of the PSF width produces the poorest results, characterized by strong overfitting and loss of quantitative accuracy. 
The best overall performance is obtained when using the inferred PSF width together with a modest overestimation of the noise level, for example \(NL = 200\%\).

\subsection{GE Discovery MI -- PSF modeling}
\label{sec:results_ge}

In clinical practice, PET reconstruction algorithms often incorporate point-spread-function (PSF) modeling to enhance spatial resolution. 
While this generally improves contrast recovery, it may also introduce artifacts such as edge overshoot or ringing, which in some cases motivates the use of reconstructions without PSF modeling \citep{rahmim2013resolution,munk2017point,tsutsui2017edge}. 
Here, we apply the MlPET method to NEMA phantom images reconstructed both with and without PSF modeling using data from the GE Discovery MI scanner, in order to evaluate the impact of PSF modeling on image quality and quantitative accuracy.

\begin{figure}
    \centering
    \includegraphics[width=1.00\textwidth]{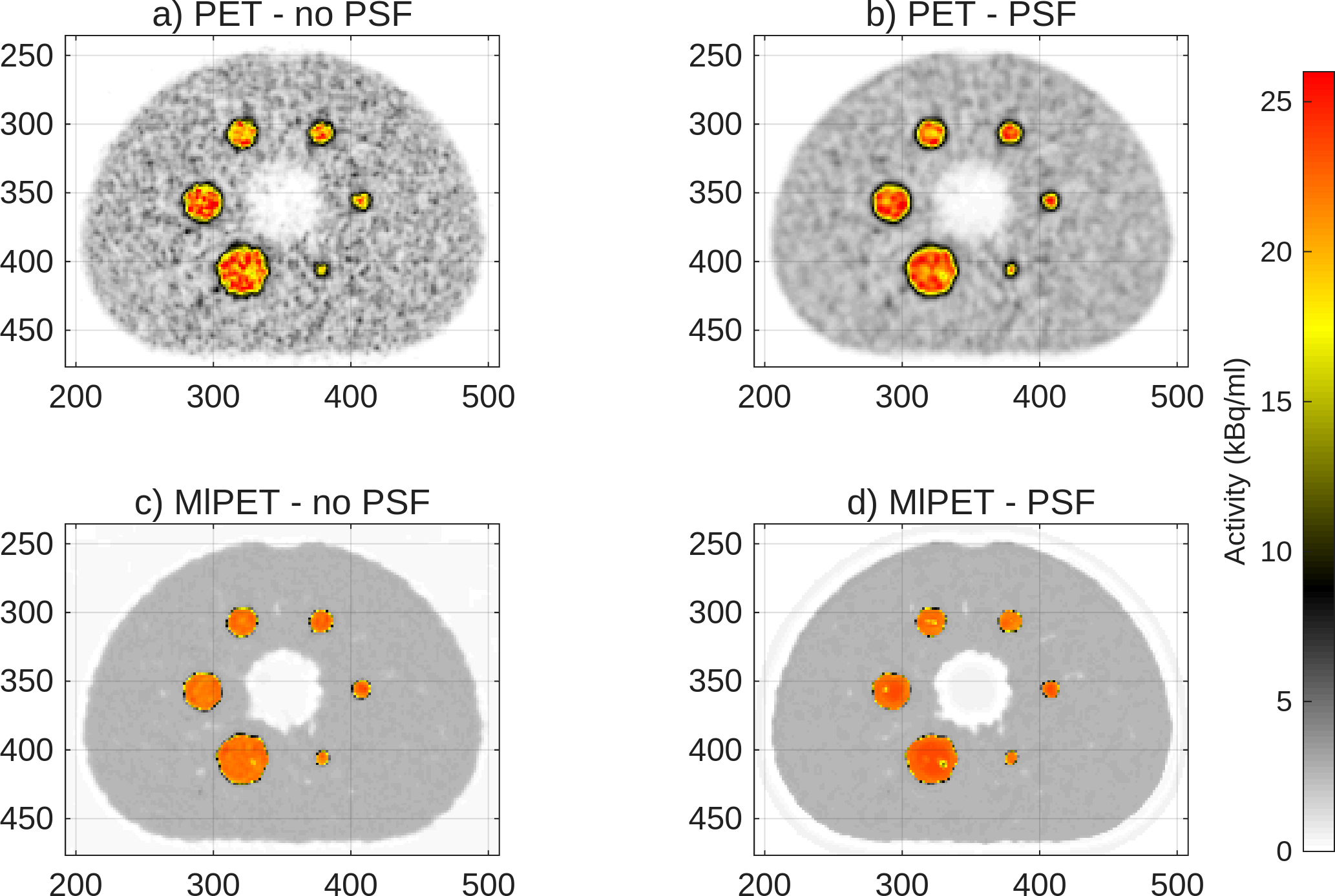}
    \caption{Two-dimensional x-slice through the centers of the six spheres in the NEMA phantom, showing standard PET reconstructions without PSF (a) and with (b) PSF modeling in the reconstruction, and the corresponding activity concentration estimated using the MlPET algorithm (c,d).}
    \label{fig:ge_slice}
\end{figure}

Figures~\ref{fig:ge_slice}a--b show transaxial slices through the centers of the six NEMA spheres obtained using standard PET reconstruction without (a) and with (b) PSF modeling. 
As expected, the reconstruction without PSF modeling (Figure~\ref{fig:ge_slice}a) exhibits substantially higher noise compared to the reconstruction with PSF modeling (Figure~\ref{fig:ge_slice}b). 
Figures~\ref{fig:ge_slice}c--d present the corresponding posterior mean activity concentrations estimated using MlPET, again applied to data reconstructed without (c) and with (d) PSF modeling. 
In both cases, MlPET markedly reduces noise relative to the standard reconstructions and provides visibly enhanced spatial resolution.

\begin{figure}
    \centering
    \includegraphics[width=1.00\textwidth]{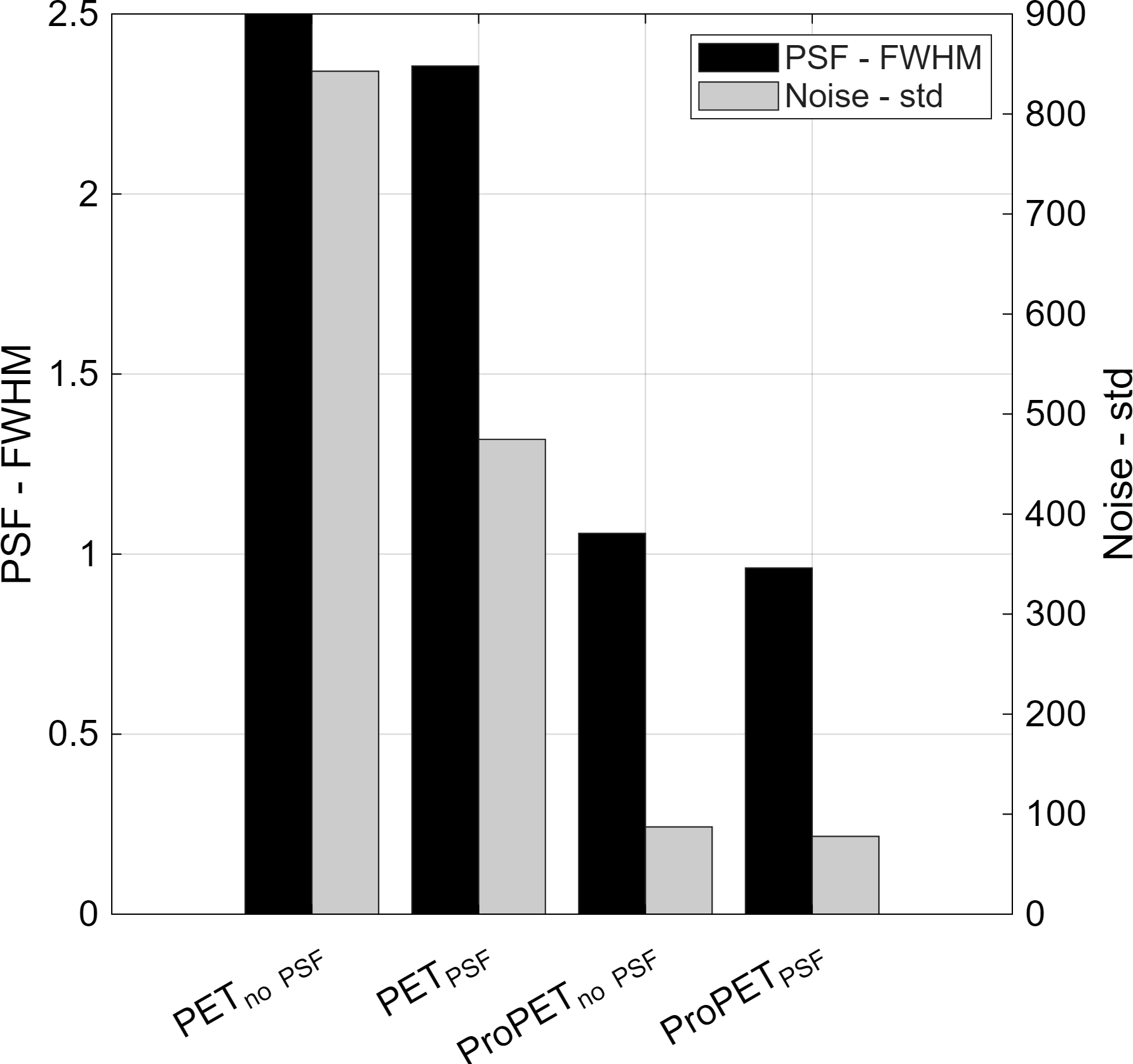}
    \caption{Estimated PSF full width at half maximum (FWHM) and standard deviation of the background signal, obtained from standard PET and MlPET reconstructions with and without PSF modeling.}
    \label{fig:ge_psf_noise}
\end{figure}

These visual observations are supported quantitatively by Figure~\ref{fig:ge_psf_noise}, which shows a clear reduction in the effective PSF width and a significant decrease in background noise, measured as the standard deviation within a uniform background region. 
Collectively, these results demonstrate that MlPET improves both spatial resolution and noise characteristics compared with conventional PET reconstructions, regardless of whether PSF modeling is included in the reconstruction process.

Figure~\ref{fig:ge_psf_noise} compares the estimated noise level in the low-activity region and the FWHM of the effective PSF obtained from PET and MlPET reconstructions, both without and with PSF modeling. 
It is clear that MlPET achieves lower noise and higher spatial resolution (smaller PSF width) than standard PET reconstruction, irrespective of whether PSF modeling is used. 
As expected, for conventional PET reconstruction, incorporating PSF modeling reduces the estimated noise level. 
In contrast, when applying MlPET, the estimated noise level is nearly identical for reconstructions with and without PSF modeling. 
This finding suggests that when MlPET is used, one can choose to disregard PSF modeling during PET reconstruction, and obtain similar results with MlPET.

\subsection{Siemens Biograph Vision Quadra – Reduced scan time}
\label{sec:results_quadra}

The Siemens Biograph Vision Quadra Edge is a long axial field-of-view (LAFOV) total-body PET/CT system that enables simultaneous imaging of large body regions with high sensitivity and spatial resolution \citep{prenosil2022performance}. 
Its extended axial field of view allows for substantially shorter scan times compared to conventional PET scanners while maintaining high image quality. 
Reducing scan duration is particularly advantageous for patients who may struggle to remain still, such as children or individuals in pain, since minimizing motion artifacts and the need for sedation improves diagnostic reliability and patient comfort. 
Several studies have demonstrated that the Quadra system can achieve significant reductions in acquisition time without compromising image quality or quantitative accuracy, thereby increasing scanner throughput and reducing radiation exposure \citep{hornnes2021effect,ingbritsen2025optimisation}.

In this study, we evaluate the performance of MlPET applied to data from the NEMA IEC body phantom acquired on the Siemens Biograph Vision Quadra Edge. 
The same data set was reconstructed using a range of acquisition durations, $t = [1, 2, 5, 10, 20, 40, 80, 160, 320, 640, 900]$~seconds, 
resulting in 11 reconstructed PET images used to assess MlPET performance for different scan times.

\subsubsection{Visual assessment}

\begin{figure}[h]
    \centering
    \includegraphics[width=1.00\textwidth]{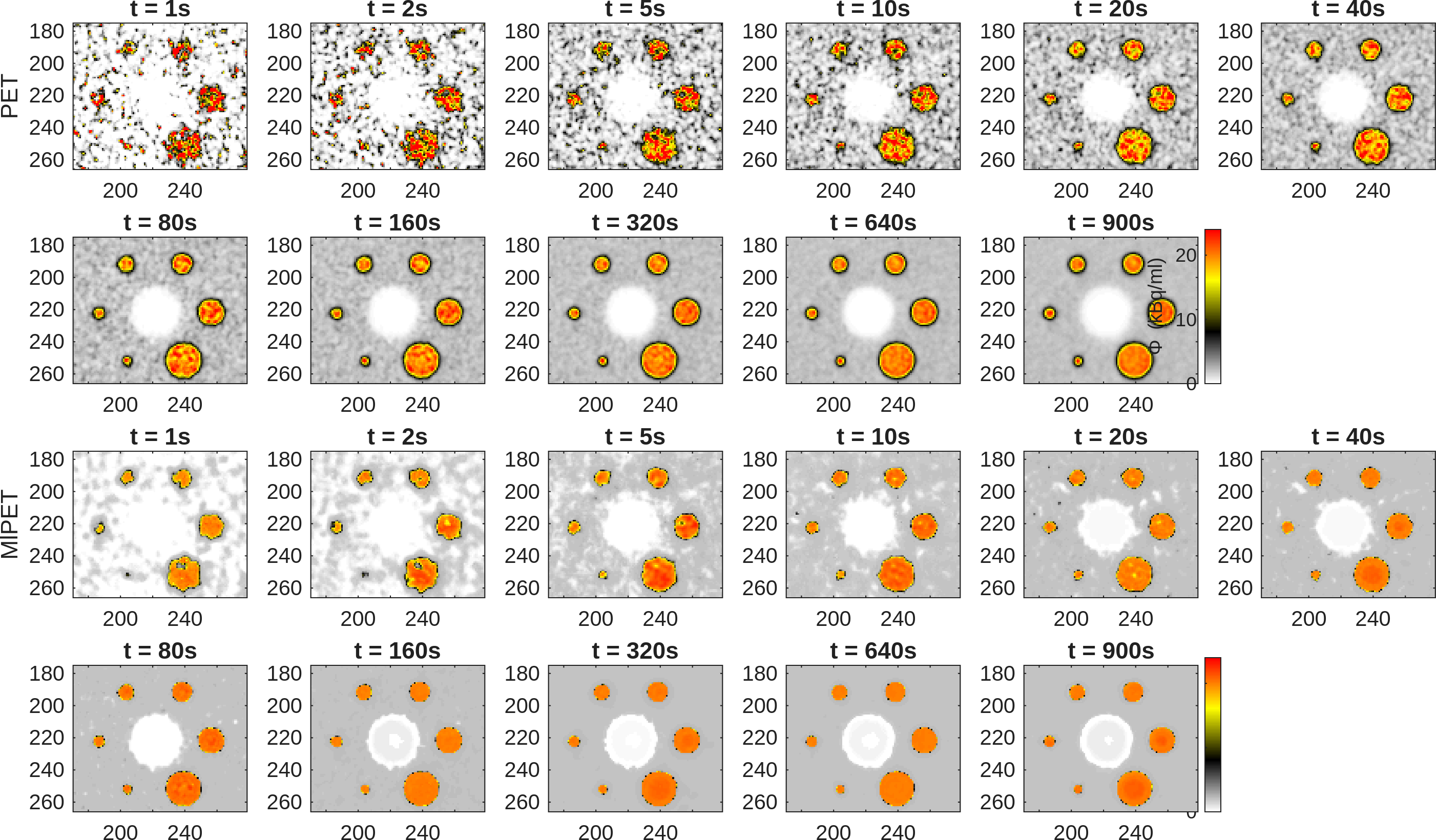}
    \caption{2D slice of the activity concentration obtained using PET (top) and MlPET(bottom) through the center of the 6 phantom spheres.}
    \label{fig:quadra_compare}
\end{figure}

Figure~\ref{fig:quadra_compare} shows 2D transaxial slices through the centers of the six NEMA spheres, comparing conventional PET (top) and MlPET (bottom) for acquisition times ranging from 1 to 900~seconds. 
MlPET demonstrates enhanced lesion boundary definition and reduced background noise compared to conventional PET, particularly at shorter acquisition durations. 
At just 5~seconds, MlPET successfully visualizes all spheres with clearly defined boundaries, whereas conventional PET requires substantially longer acquisitions to achieve comparable visual quality.

\begin{figure}[h]
\centering
\includegraphics[width=0.75\textwidth]{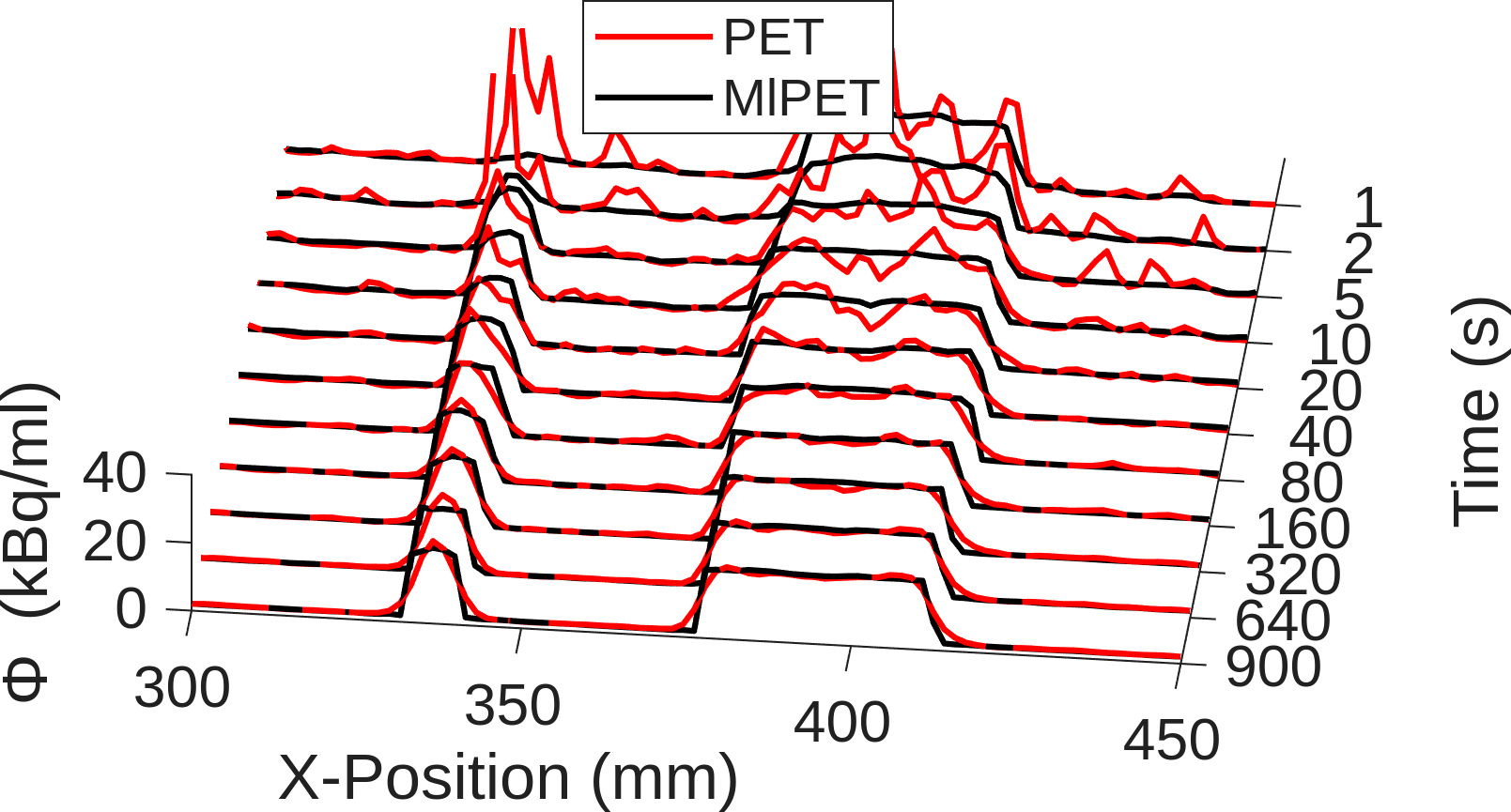}
\caption{Profile through the center of the smallest and largest phantom region, comparing PET (red) and MlPET (black).}
\label{fig:quadra_wfall}
\end{figure}

Figure~\ref{fig:quadra_wfall} presents activity concentration profiles across the smallest (10\,mm) and largest (37\,mm) spheres. 
These profiles confirm that MlPET (black) provides noticeably sharper lesion boundaries and lower background noise compared to conventional PET (red), highlighting the method’s ability to maintain high resolution even at very short acquisition times.

\subsubsection{Quantitative analysis}

\begin{figure}[h]
    \centering
    \includegraphics[width=0.75\textwidth]{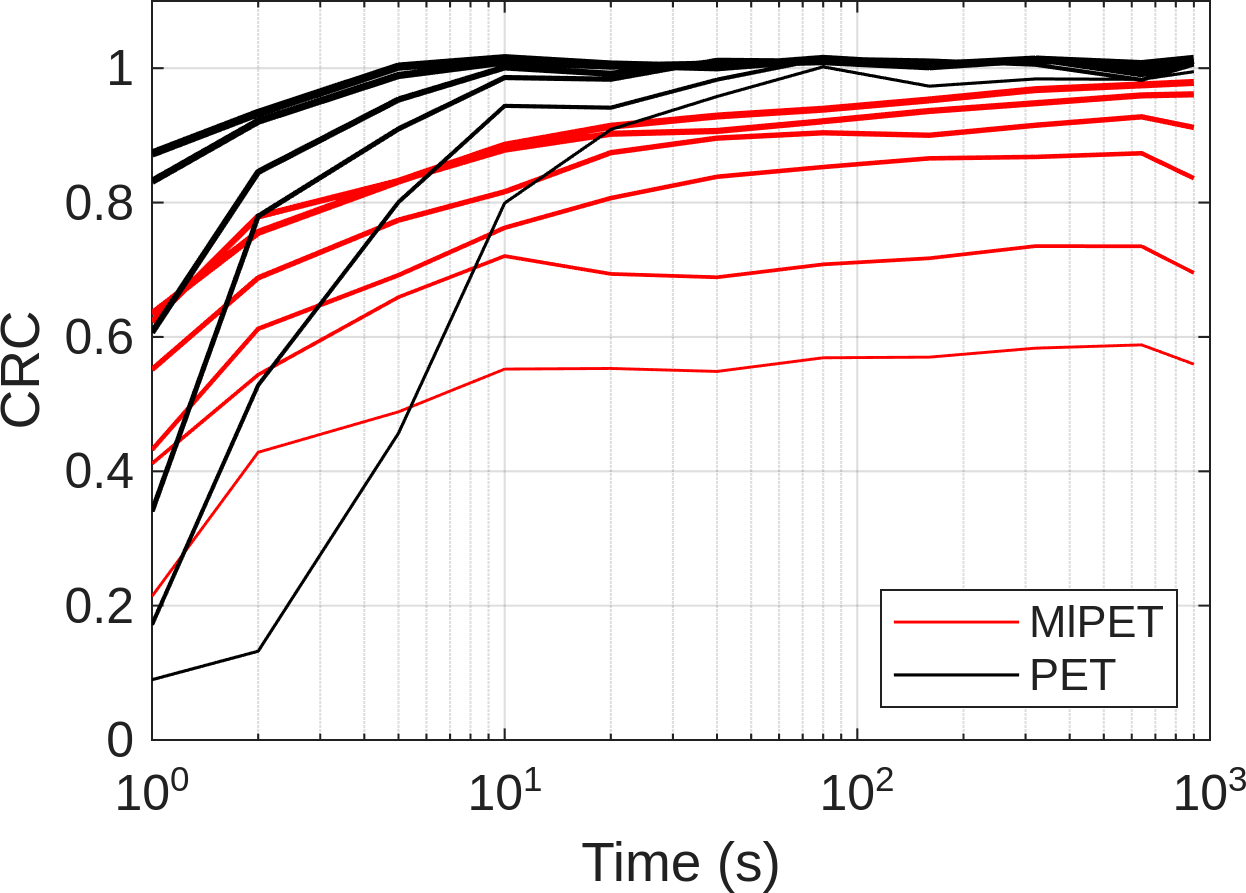}
    \caption{Median contrast recovery coefficient (CRC, Eqn. \ref{eqn:CRC},) for spheres of different sizes (thinner lines represent smaller spheres) for MlPET (black) and PET (red)}\label{fig:quadra_crc}
\end{figure}

Figure~\ref{fig:quadra_crc} shows the Contrast Recovery Coefficient (CRC) values across 11 acquisition durations and six sphere sizes. The CRC are also listen in Tables~\ref{tab:crc_quadra_pet} and~\ref{tab:crc_quadra_propet}. 
MlPET achieves CRC values approaching unity for all sphere diameters at scan times of 40~seconds or longer, including the smallest 10\,mm sphere (CRC = 0.99 at 900~seconds), whereas conventional PET reaches only 0.56 for the same sphere. 

Importantly, MlPET attains high contrast recovery at substantially shorter acquisition durations, with 40–80~second scans yielding CRC values that exceed those of conventional PET even at 900~seconds.

The evaluation using phantom data from the Siemens Biograph Vision Quadra demonstrates that MlPET provides improved contrast recovery, reduced noise, and enhanced spatial resolution compared to conventional PET reconstruction when applied to the NEMA IEC body phantom. 
In this controlled setting, MlPET with 40–80~second acquisitions achieved image quality metrics comparable to or exceeding those of conventional PET at 900~seconds, indicating the potential for a substantial reduction in required scan time.

These results, however, are based solely on phantom data acquired under idealized conditions. 
The NEMA IEC body phantom is a simplified model with known geometry, uniform background, and spherical lesions of well-defined activity ratios. 
Clinical PET imaging presents additional challenges, including irregular lesion shapes, heterogeneous backgrounds, patient motion, variable tracer uptake, and anatomical variability—factors not represented in phantom experiments. 
Further validation using patient data is therefore essential to determine whether the observed performance improvements translate to clinical settings and to assess the robustness of the method across diverse patient populations and pathological conditions before any conclusions can be drawn about clinical applicability.

\subsection{Clinical example – Breast cancer patient}

\begin{figure}[h]
    \centering
    \includegraphics[width=0.99\textwidth]{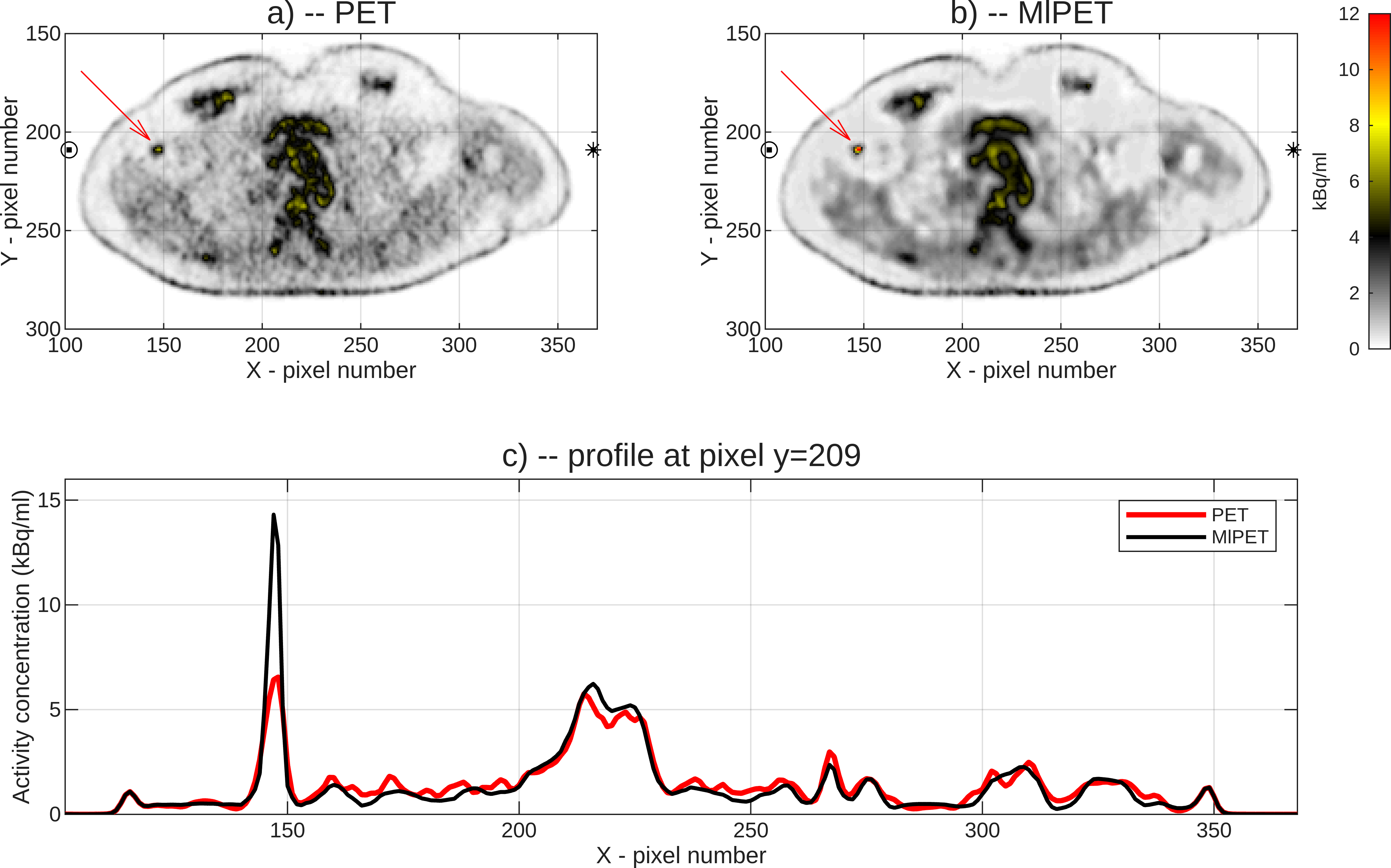}
    \caption{MlPET applied to clinical data from a breast cancer patient imaged with 18F-FDG PET/CT. a,b) 2D transaxial slice number 349 using traditional PET (a)  and MlPET (b). c) X-profile through a parasternal lymph node (as indicated by the red arrow in a) and b).}
    \label{fig:clinical}
\end{figure}

As a clinical example, we consider a breast cancer patient imaged with \textsuperscript{18}F-FDG PET/CT. 
Figure~\ref{fig:clinical}a shows a transaxial PET slice reconstructed using standard OSEM, while Figure~\ref{fig:clinical}b displays the corresponding posterior mean activity concentration estimated using MlPET with the prior model \(\rho_c(\mathbf{\Phi})\) described in Section~\ref{sec:prior}. 
The data were acquired on a Siemens Biograph Vision 600 scanner using the same reconstruction protocol as in Section~\ref{sec:results_600}, with an acquisition time of 300~seconds.

Figure~\ref{fig:clinical}c presents intensity profiles through the parasternal lymph node along the x-axis, as indicated in Figures~\ref{fig:clinical}a–b, comparing the standard PET reconstruction (red) and the MlPET result (black). 

In the region surrounding the parasternal lymph node, the variability in activity is visibly reduced when using MlPET. 
Moreover, the spatial resolution and contrast ratio appear improved at the lymph node location, suggesting that the true activity concentration may be higher than indicated by the conventional PET image. 
This implies enhanced detectability of the small lesion associated with the lymph node.

Overall, this single clinical example indicates improved sensitivity and reduced image noise when using MlPET compared to standard PET reconstruction, consistent with the findings from the phantom analysis. 
Nevertheless, a larger study including diverse patient data and appropriate quantitative validation is required before definitive conclusions can be drawn regarding the clinical applicability of the method.

\section{Discussion}\label{sec:discussion}

\subsection{Relation to prior work}
MlPET builds directly upon the probabilistic inversion framework introduced by \citet{hansen2023probabilistic}, which formulated PET image analysis as a probabilistic inverse problem incorporating informed priors, PSF modeling, and correlated noise. 
While that approach demonstrated the feasibility of probabilistic PET deconvolution, it was computationally constrained by Markov chain Monte Carlo sampling. 
In parallel, \citet{hansen2022use} proposed a neural strategy for estimating posterior statistics in general inverse problems without explicit sampling. 
MlPET combines these concepts, retaining the interpretability and rigor of the probabilistic formulation while enabling efficient voxelwise inference across entire 3D PET volumes. 
This integration transforms probabilistic PET analysis from a theoretical framework into a practical methodology for routine post-reconstruction evaluation.

\subsection{Computational performance and practicality}
The MlPET workflow is computationally efficient, allowing practical post-reconstruction analysis of full 3D PET datasets. 
A typical training dataset consisted of $N = 8\times10^5$ localized examples. 
Generating prior realizations and simulating corresponding noise required about 8~minutes in total, and network training an additional 4~minutes, repeated three times to ensure convergence. 
Application of the trained network to a full 3D PET volume took approximately 4~minutes on a standard workstation, resulting in an overall runtime below 20~minutes.

Compared to the original probabilistic deconvolution framework (ProPET) \cite{hansen2023probabilistic}, which relied on explicit posterior sampling using the extended Metropolis algorithm, MlPET provides a speed-up of several orders of magnitude. 
In ProPET, sampling even a small 2D subset (approximately $90\times90$ voxels) required several hours, while extending to full 3D volumes was computationally prohibitive. 
The enormous memory demands and high dimensionality of the posterior space make such random walks infeasible, as obtaining statistically independent samples would require an unmanageable number of iterations. 
By replacing explicit sampling with localized neural network, MlPET overcomes these limitations and enables efficient estimation of posterior statistics for large-scale PET data.

\subsection{Reduced scan time}
Several commercial and research methods have been developed to enable reduced-count or short-duration PET imaging. 
Deep learning approaches such as SubtlePET and related CNN- or GAN-based techniques have demonstrated effective noise suppression at 30--50\% of standard counts \cite{bonardel2022clinical,schaefferkoetter2020convolutional,wang20183d,chen2019ultra,kruzhilov2024whole}. 
However, these methods typically rely on large scanner-specific paired datasets, which limits their generalizability \cite{gong2019pet}. 
In contrast, MlPET achieves comparable improvements in image quality without retraining for individual scanners or protocols and, unlike purely data-driven methods, simultaneously enhances both noise and resolution through explicit probabilistic and physical modeling. 
This capability suggests that substantial reductions in acquisition time are achievable while maintaining diagnostic image quality.

\subsection{Clinical applicability and future perspectives}
The present evaluation was based primarily on controlled phantom experiments, enabling quantitative benchmarking of contrast recovery, noise, and resolution. 
A single clinical example was also included to demonstrate the potential applicability of MlPET to patient data. 
In this case, the method produced qualitatively similar improvements to those seen in the phantom studies, showing visibly reduced noise and enhanced lesion contrast. 
While this result is encouraging, it represents only an initial demonstration. 
For evaluation of the diagnostic potential further studies are required
In real-world imaging, factors such as heterogeneous tissue composition, irregular lesion morphology, tracer kinetics, and patient motion will influence both the data and the suitability of the chosen prior model. 
A key strength of MlPET is its ability to incorporate informed priors that reflect domain-specific physiological knowledge. 
In many applications—particularly brain imaging—such priors can be derived from anatomical atlases, regional uptake patterns, or kinetic models. 
Future work will focus on developing and validating such priors for clinical use, enabling robust probabilistic PET analysis across a wide range of imaging contexts.

\section{Conclusion}\label{sec13}

We have presented MlPET, a fast, localized machine learning framework for probabilistic post-reconstruction analysis of PET images that addresses key limitations of conventional PET imaging. 
By replacing computationally intensive sampling with an efficient neural network–based approximation, MlPET enables practical application of probabilistic analysis to full 3D PET volumes while explicitly incorporating prior information on tissue activity distributions, point-spread-function characteristics, and correlated noise.

In controlled phantom studies across three PET systems (Siemens Biograph Vision Quadra, Siemens Biograph Vision 600, and GE Discovery MI), MlPET yielded substantial improvements in key imaging metrics. Contrast recovery coefficients (CRCs) moved consistently toward unity across sphere sizes and surpassed those of standard PET, with several conditions producing values near 1.0 for the 10 mm sphere. The estimated effective PSF full width at half maximum (FWHM) was markedly reduced, decreasing from above 2 mm in standard PET to below 1 mm with MlPET,
corresponding to roughly a 2.5× reduction in the effective point-spread width.
Notably, MlPET images acquired with 40–80~s scan times exhibited image quality comparable to, or better than, conventional PET images acquired at 900~s.

These findings represent a proof-of-concept demonstration under idealized conditions using the NEMA IEC body phantom, which provides well-defined geometry, uniform backgrounds, and known activity distributions. 
This controlled setup enabled quantitative benchmarking of the method’s theoretical performance. 
However, clinical PET imaging introduces additional complexities, including heterogeneous tissue backgrounds, irregular lesion shapes, patient motion, variable tracer kinetics, and anatomical variability not captured in phantom experiments.

The sensitivity analysis showed that MlPET remains robust across a range of noise model parameters, with optimal results achieved when slightly overestimating the noise level to account for uncertainties in the forward model. 
This robustness to model mismatch is an encouraging indicator for clinical translation, where precise noise characterization may be challenging.

Future work will focus on clinical validation of MlPET using patient data to determine whether the improvements observed in phantom studies translate into diagnostic benefit. 
Important directions include developing appropriate prior models for different clinical contexts, assessing robustness across diverse patient populations and imaging protocols, and exploring optimal integration of anatomical information from complementary modalities. 
Prospective clinical studies will also be necessary to confirm whether the potential for reduced scan times can be realized without compromising diagnostic accuracy or lesion detectability.

Overall, MlPET represents a promising advancement in PET image analysis that leverages machine learning to achieve computationally efficient probabilistic inference while maintaining the interpretability and theoretical rigor of probabilistic inverse methods. 
If validated in clinical settings, MlPET could contribute to improved detection of small lesions, reduced radiation exposure through shorter acquisitions, and enhanced patient comfort, all while maintaining or improving diagnostic image quality.

\backmatter

\bmhead{Acknowledgements}
Phantom data have been acquired at the Department of Nuclear Medicine and PET Centre, Aarhus University Hospital, Denmark.

\section*{Declarations}

\begin{itemize}
\item This work has been partly funded by a grant from BetaHealth (\#XXXX).
\item Thomas Mejer Hansen is the CEO of NoloSight, a company developing AI-based solutions for medical imaging.
\end{itemize}

\clearpage
\begin{appendices}

\section{Noise and PSF inference}
Table~\ref{tab:inferred_noise_params} summarizes the inferred noise model parameters 
($a$ and $b$, representing the power-law model relating the standard deviation to the mean activity) 
along with the correlation range (FWHM, $r$) for the noise and the effective PSF range (FWHM, $r_{\mathrm{psf}}$) 
for the GE Discovery MI (with and without PSF modeling), Siemens Biograph Vision 600, 
and Siemens Biograph Vision Quadra scanners.
\begin{table}[h]
    \centering
    \caption{Inferred noise model parameters for different scanners.}
    \label{tab:inferred_noise_params}
    \begin{tabular}{|l|c|c|c|c|}
    \hline
     Scanner & a & b & r (mm) & $r_{psf} (mm)$ \\ \hline
     GE Discovery MI (PSF) & 0.26 & 0.63 & 7.5 & 2.4 \\ \hline
     GE Discovery MI (No PSF) & 0.48 & 0.57 & 5.3 & 2.5 \\ \hline
     Siemens Vision 600 & 0.22 & 0.57 & 6.1 & 2.4 \\ \hline
    Quadra$_{t=1\text{s}}$ & 3.48 & 0.47 & 3.5 & 1.4 \\ \hline
    Quadra$_{t=2\text{s}}$ & 2.41 & 0.48 & 4.0 & 1.6 \\ \hline
    Quadra$_{t=5\text{s}}$ & 1.19 & 0.59 & 4.6 & 1.8 \\ \hline
    Quadra$_{t=10\text{s}}$ & 0.81 & 0.65 & 4.5 & 1.9 \\ \hline
    Quadra$_{t=20\text{s}}$ & 0.57 & 0.63 & 4.6 & 2.0 \\ \hline
    Quadra$_{t=40\text{s}}$ & 0.40 & 0.64 & 4.8 & 2.0 \\ \hline
    Quadra$_{t=80\text{s}}$ & 0.27 & 0.70 & 4.9 & 2.1 \\ \hline
    Quadra$_{t=160\text{s}}$ & 0.19 & 0.69 & 4.9 & 2.0 \\ \hline
    Quadra$_{t=320\text{s}}$ & 0.14 & 0.66 & 4.9 & 2.0 \\ \hline
    Quadra$_{t=640\text{s}}$ & 0.10 & 0.68 & 4.9 & 2.0 \\ \hline
    Quadra$_{t=900\text{s}}$ & 0.08 & 0.68 & 5.5 & 2.2 \\ \hline
    \end{tabular}
\end{table}

\section{Siemens Quadra}
Tables~\ref{tab:crc_quadra_pet} and~\ref{tab:crc_quadra_propet} present the median Contrast Recovery Coefficient ($\mathrm{CRC}_{\mathrm{median}}$) values for the Siemens Biograph Vision Quadra, obtained from the standard PET reconstruction and after applying MlPET, respectively, across different sphere sizes and acquisition times.
\begin{table}
    \centering
    \caption{$CRC_{median}$ - Siemens Quadra PET}
    \label{tab:crc_quadra_pet}
    \begin{tabular}{|c|c|c|c|c|c|c|c|c|c|c|c|}
    \hline
    Size (mm) & \multicolumn{11}{c|}{Time (s)} \\ \hline
     & 1 & 2 & 5 & 10 & 20 & 40 & 80 & 160 & 320 & 640 & 900 \\ \hline
    10 mm & 0.21 & 0.43 & 0.49 & 0.55 & 0.55 & 0.55 & 0.57 & 0.57 & 0.58 & 0.59 & 0.56 \\ \hline
    13 mm & 0.41 & 0.54 & 0.66 & 0.72 & 0.69 & 0.69 & 0.71 & 0.72 & 0.74 & 0.73 & 0.70 \\ \hline
    17 mm & 0.43 & 0.61 & 0.69 & 0.76 & 0.81 & 0.84 & 0.85 & 0.87 & 0.87 & 0.87 & 0.84 \\ \hline
    22 mm & 0.55 & 0.69 & 0.77 & 0.82 & 0.87 & 0.90 & 0.90 & 0.90 & \cellcolor{gray!20}0.92 & \cellcolor{gray!20}0.93 & \cellcolor{gray!20}0.91 \\ \hline
    28 mm & 0.62 & 0.78 & 0.83 & 0.88 & 0.90 & \cellcolor{gray!20}0.91 & \cellcolor{gray!20}0.92 & \cellcolor{gray!20}0.94 & \cellcolor{gray!20}0.95 & \cellcolor{gray!20}0.96 & \cellcolor{gray!20}0.96 \\ \hline
    37 mm & 0.63 & 0.76 & 0.83 & 0.89 & \cellcolor{gray!20}0.91 & \cellcolor{gray!20}0.93 & \cellcolor{gray!20}0.94 & \cellcolor{gray!20}0.95 & \cellcolor{gray!20}0.97 & \cellcolor{gray!20}0.98 & \cellcolor{gray!20}0.98 \\ \hline
    \end{tabular}
\end{table}
    
\begin{table}
    \centering
    \caption{$CRC_{median}$ - MlPET on Siemens Quadra PET}
    \label{tab:crc_quadra_propet}
    \begin{tabular}{|c|c|c|c|c|c|c|c|c|c|c|c|}
    \hline
    Size (mm) & \multicolumn{11}{c|}{Time (s)} \\ \hline
    & 1 & 2 & 5 & 10 & 20 & 40 & 80 & 160 & 320 & 640 & 900 \\ \hline
    10 mm & 0.09 & 0.13 & 0.46 & 0.80 & \cellcolor{gray!20}0.91 & \cellcolor{gray!20}0.96 & \cellcolor{gray!20}1.00 & \cellcolor{gray!20}0.97 & \cellcolor{gray!20}0.98 & \cellcolor{gray!20}0.98 & \cellcolor{gray!20}0.99 \\ \hline
    13 mm & 0.17 & 0.53 & 0.80 & \cellcolor{gray!20}0.94 & \cellcolor{gray!20}0.94 & \cellcolor{gray!20}0.98 & \cellcolor{gray!20}1.01 & \cellcolor{gray!20}1.01 & \cellcolor{gray!20}1.01 & \cellcolor{gray!20}0.98 & \cellcolor{gray!20}1.01 \\ \hline
    17 mm & 0.34 & 0.78 & \cellcolor{gray!20}0.91 & \cellcolor{gray!20}0.99 & \cellcolor{gray!20}0.98 & \cellcolor{gray!20}1.01 & \cellcolor{gray!20}1.01 & \cellcolor{gray!20}1.01 & \cellcolor{gray!20}1.01 & \cellcolor{gray!20}0.99 & \cellcolor{gray!20}1.01 \\ \hline
    22 mm & 0.61 & 0.85 & \cellcolor{gray!20}0.95 & \cellcolor{gray!20}1.00 & \cellcolor{gray!20}0.99 & \cellcolor{gray!20}1.01 & \cellcolor{gray!20}1.01 & \cellcolor{gray!20}1.01 & \cellcolor{gray!20}1.01 & \cellcolor{gray!20}1.00 & \cellcolor{gray!20}1.01 \\ \hline
    28 mm & 0.83 & \cellcolor{gray!20}0.92 & \cellcolor{gray!20}0.99 & \cellcolor{gray!20}1.01 & \cellcolor{gray!20}1.00 & \cellcolor{gray!20}1.01 & \cellcolor{gray!20}1.01 & \cellcolor{gray!20}1.00 & \cellcolor{gray!20}1.01 & \cellcolor{gray!20}1.00 & \cellcolor{gray!20}1.01 \\ \hline
    37 mm & 0.87 & \cellcolor{gray!20}0.93 & \cellcolor{gray!20}1.00 & \cellcolor{gray!20}1.01 & \cellcolor{gray!20}1.01 & \cellcolor{gray!20}1.00 & \cellcolor{gray!20}1.01 & \cellcolor{gray!20}1.00 & \cellcolor{gray!20}1.01 & \cellcolor{gray!20}1.01 & \cellcolor{gray!20}1.01 \\ \hline
    \end{tabular}

\end{table}

\end{appendices}

\clearpage
\bibliography{refs}

\end{document}